\documentclass[aps,prx,superscriptaddress,longbibliography,twocolumn]{revtex4-2}

\pdfoutput=1

\usepackage[toc,page,title,titletoc,header]{appendix}
\usepackage{amssymb}
\usepackage{amsmath}
\usepackage{graphicx,bm}
\usepackage{epstopdf}
\usepackage{subfigure}
\usepackage{natbib}
\usepackage{epsfig}
\usepackage{amsfonts}
\usepackage{mathrsfs}
\usepackage{CJK}
\usepackage{array,longtable}
\usepackage{float}
\usepackage[usenames, dvipsnames, x11names]{xcolor}
\usepackage[colorlinks]{hyperref}
\usepackage[capitalise]{cleveref}
\hypersetup{
 colorlinks=true,
 citecolor=blue,
 linkcolor=blue,
 urlcolor=cyan}
\usepackage{mathtools}

\usepackage{bbold}
\usepackage{natbib}
\usepackage{svg}
\usepackage{float}
\usepackage{bbm}
\usepackage{lipsum}

\newcommand{\tr}{ \ \mathrm{tr}}
\newcommand{\R}{\mathbb{R}}

\newcommand{\dd}{\ \mathrm{d}}

\newcommand{\Z}{\mathbb{Z}}

\newcommand{\D}{\mathcal{D}}

\newcommand{\Pf}{\mathrm{Pf}}

\DeclareMathOperator{\hc}{H.c.}

\newcommand{\arccot}{\mathrm{arccot}}

\newcommand{\floor}[1]{\lfloor #1 \rfloor}

\begin{document}

\title{Insights into decohered critical states using an exact solution to matchgate circuits with Pauli noise}

\author{Andrew Pocklington}
\email{abpocklington@uchicago.edu}
\affiliation{Department of Physics, University of Chicago, 5640 South Ellis Avenue, Chicago, Illinois 60637, USA }

\affiliation{Pritzker School of Molecular Engineering, University of Chicago, Chicago, IL 60637, USA}

\author{Aashish A. Clerk}
\email{aaclerk@uchicago.edu}
\affiliation{Pritzker School of Molecular Engineering, University of Chicago, Chicago, IL 60637, USA}

\date{\today}

\begin{abstract}
The fate of non-trivial many-body states subject to decoherence is of both fundamental and practical interest.  Here, we demonstrate a new analytic technique that allows for an exact treatment of dynamics of observables in matchgate circuits subject to arbitrary Pauli noise.  We use this to obtain new insights on how decoherence influences critical ground states, focusing on the 1D transverse field Ising model subject to local Markovian Pauli noise. While such noise cannot kill the critical behavior of spin correlation functions, we show that it does lead to a surprising non-equilibrium state, with experimental signatures that are measurable without requiring post-selection or multiple copies of the system.  Despite the infinite-temperature nature of the dissipation, the decohered state is characterized by a thermal distribution of low-energy quasi-particles.  This is the direct consequence of a noise-induced emergent length scale that manifests itself in fermionic correlators.  We show how these phenomena are directly accessible in experiments using a single probe qubit, and that our results also hold for a different dephased critical state (that of an XX spin chain in the zero magnetization sector).  
\end{abstract}

\maketitle

{\section{Introduction}}

State-of-the-art quantum simulators 
\cite{Leseleuc2019,Semeghini2021,Sompet2022,fang2025,su2025,sun2026}
and quantum processors 
\cite{Zhu2020,Satzinger2021,Haghshenas2024,Mi2024,Evered2025}
have made rapid progress in their ability to prepare complex quantum states.  This include states with topological order and long-range entanglement \cite{Leseleuc2019,Semeghini2021,Satzinger2021,Sompet2022,su2025,Evered2025}, and scale-invariant quantum critical states which appear at quantum phase transitions \cite{Zhu2020,Haghshenas2024,Mi2024,fang2025,sun2026}. Unavoidable noise and dissipation imply that the dynamics of these systems is necessarily non-unitary, motivating a host of new fundamental questions concerning how novel quantum states behave in out-of-equilibrium settings, and the basic nature of mixed state phases of matter \cite{Sang2024,Rakovszky2024,Sang2025,Wang2025b,Sohal2025,Ellison2025}. Recent works have studied the robustness of topologically ordered states to noise \cite{bao2023,Fan2024,Ellison2025,Sohal2025,Sala2025,wang2025}, as well as the effect of non-unitary dynamics on critical ground states \cite{Lee2023,Lin2023,Zou2023,Garratt2023,Murciano2023,Sala2024,khanna2025}.  While these works reveal a host of interesting physics, they have primarily studied quantities that are well-suited to theoretical techniques like CFT, but that are extremely challenging for experiments. These include quantities nonlinear in the system density matrix (which require multiple copies of the system to measure) \cite{Zou2023}, and quantities post-selected on measurement outcomes \cite{Lin2023,Garratt2023,Murciano2023} (which generally incur an exponential sampling overhead).  It would be extremely interesting to study how non-unitary evolution of non-trivial topological or critical states impacts simple observable quantities, and what insights these provide into the nature of the resulting non-equilibrium state.  

\begin{figure}[t!]
    \centering
    \includegraphics[width = \linewidth]{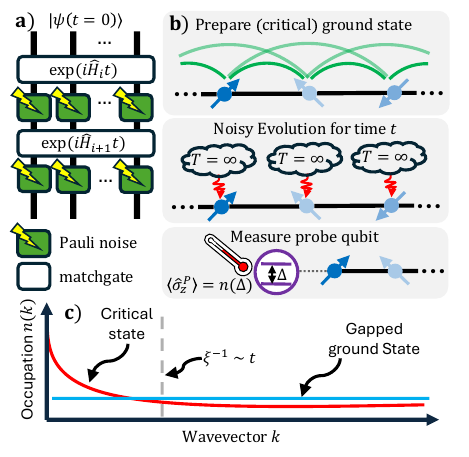}
    \caption{a) Schematic of a matchgate circuit with Pauli noise.  Despite non-trivial fermionic strings, such circuits can be efficiently simulated using our technique. b) Experimental protocol where one 
    first prepares a critical ground state, then decoheres the state under noisy evolution, then finally uses a probe qubit with tunable splitting $\Delta$ to measure the frequency-dependent effective temperature. c) Example quasiparticle excitation density as a function of wavevector for either a decohered critical ground state (red) or gapped ground state (blue). The critical state roughly follows a Fermi-Dirac distribution up to a length scale $\xi^{-1}$ which grows linearly with the total time of the decoherence process as depicted in b).  }
    \label{fig:cartoon}
\end{figure}

In this work, we address these questions through a detailed study of the critical ground state of the 1D transverse-field Ising model (TFIM) subject to local, Markovian Pauli noise.  Our focus is on unconditional evolution (no post-selection), and on standard observables that are linear in the density matrix.  One might immediately fear that in this setting, nothing interesting can happen, as it is well-known that local noise channels cannot change critical properties: spin correlation functions must decay with the same power law as they did without the application of dissipation.  Surprisingly, we show that this is not the full story.  While spin correlators have an unchanged algebraic decay, we show that noise causes fermionic correlators (which are also critical in the ground state) to immediately develop a length scale (i.e.~they now decay exponentially).  Hence, after decoherence, the spins remain critical, whereas the fermions do not. 

We further show that this dichotomy has measurable consequences:  it leads to a novel non-equilibrium state whose low-energy physics is described by an effective time-dependent temperature.  This is surprising, as Markovian Pauli noise corresponds to an infinite-temperature, infinite-bandwidth environment; such environments typically do not lead to finite-temperature thermal equilibrium states.  The emergent temperature we find is directly tied to criticality: it does not arise if one starts with a non-critical ground state.  We also show that this effective temperature can be easily probed experimentally, using a single frequency-tunable probe qubit that is weakly coupled to the chain  (see \cref{fig:cartoon}). While our focus here is on the TFIM, we suspect that the basic mechanism we describe could apply to more general critical ground states exposed to decoherence, thus representing a new kind of emergent non-equilibrium state. To support this claim, we show that all of the features identified in the TFIM are also present in a critical XX chain in the zero magnetization sector.

Our results rely on a technical innovation that has implications well beyond our decohered TFIM model.  The questions we are interested in are not naturally suited to the CFT techniques in, e.g., Ref.~\cite{Zou2023}; also, naive master equation simulations are not feasible, as the dynamics when written in terms of Jordan-Wigner (JW) fermions now has non-trivial strings.  We circumvent these issues via a new exact solution to the dynamics of matchgate circuits \cite{Terhal2002,jozsa2008} subject to arbitrary Pauli noise, see \cref{fig:cartoon}. This builds on the results of Ref.~\cite{Pocklington2025}, which introduced a new stochastic averaging technique for a class of dissipative spin chain models.  Here, we go further, and show that for Pauli noise, one can obtain closed-form solutions for the time-evolution of arbitrary observables.  Our results here expand the class of previously-known easy-to-simulate fermionic dynamics \cite{Shtanko2021}, and also introduce a new class of efficiently simulable noisy quantum circuits.  It also allows one to go beyond the standard setting of decohered topological or critical states, and consider situations where both the decoherence and Hamiltonian evolution are simultaneously acting on the system.    

The remainder of this manuscript is organized as follows: In \cref{sec:solution}, we present the exact solution for matchgate circuits with Pauli noise. In \cref{sec:IsingModel}, we introduce the specific model we will consider, namely an Ising model with decoherence in the $X$ and $Y$ directions. In \cref{sec:CritPhenom}, we identify the physical phenomena that result from applying the decoherence channel to the Ising critical state when there are no coherent dynamics. Further, we propose a simple experiment that would be able to observe the effects. In \cref{sec:NonCrit}, we compare the results found in \cref{sec:CritPhenom} to the case where the initial condition is shifted away from the Ising critical point (but still with no coherent dynamics). In \cref{sec:WithHam}, we explain how to compute the exact solution for the dynamics in the presence of a Hamiltonian with periodic boundary conditions, and use this to show data when there are both coherent and incoherent dynamics. In \cref{sec:otherNoise}, we compare our results with other common noise channels. Finally, in \cref{sec:critXX}, we compare the results from the decohered critical TFIM to a decohered critical XX model.

\bigskip
\section{Exact Solution for matchgate Circuits with Pauli Noise}
\label{sec:solution}

We begin by showing that, for any matchgate circuit with Pauli noise, the evolution of observables can be computed exactly and efficiently. To fix terminology, we will define a unitary matchgate circuit as a circuit where every unitary operator can be understood as the free evolution of a collection of fermionic modes \cite{Terhal2002,jozsa2008}. 
In other words, there exists a mapping from qubits to free fermions such that every unitary 
maps Gaussian fermionic states to Gaussian fermionic states. For example, any unitary operator $\hat U = \exp(i \theta \hat H)$ acting on a set of spin-$1/2$'s generated by a Hamiltonian $\hat H$ of the form
\begin{align}
    \hat H &= \sum_i J_i^1 \hat \sigma_i^+ \hat \sigma_{i + 1}^- + J_i^2 \hat \sigma_i^+ \hat \sigma_{i + 1}^+ + \Delta_i \hat \sigma_i^z + \hc \label{eqn:Ham}
\end{align}
can be written exactly as a quadratic free fermion model using a Jordan-Wigner (JW) transformation \cite{Jordan1928}. If we define a set of canonical fermion modes $\hat c_i$ via
\begin{align}
    \hat \sigma_i^- &= \left( \prod_{j < i} (-1)^{\hat c_j^\dagger \hat c_j } \right) \hat c_i,
\end{align}
then we can rewrite the Hamiltonian as
\begin{align}
    \hat H &= \sum_i J_i^1 \hat c_i^\dagger \hat c_{i + 1} + J_i^2 \hat c_i^\dagger \hat c_{i + 1}^\dagger + 2 \Delta_i \hat c_i^\dagger \hat c_i + \hc \label{eqn:Ham2}
\end{align}

We are interested in studying matchgate models subject to additional non-unitary processes that correspond to Pauli noise (including cases with higher-weight Pauli operators that describe correlated noise).  Returning to a discrete time picture, this general class of problems would have evolution in each time-step (i.e.~each layer) described by e.g.~$\hat\rho(t + \delta t) = \hat U \mathcal{K}[\hat \rho] \hat U^\dagger$ where $\hat U = \exp(i \hat H \delta t)$ denotes the unitary matchgate evolution, and the noise channel $\mathcal{K}$ is defined by
\begin{align}
    \mathcal{K}[\hat \rho] &= \sum_{s \in \{0,1\}^{2n}} p_s \hat K_s \hat \rho \hat K_s^\dagger,  \\
    \hat K_s &= \prod_{i = 0}^{n-1} (\hat \sigma_i^x)^{s_{2i}} (\hat \sigma_i^z)^{s_{2i + 1}}, \label{eqn:Kraus_op}
\end{align}
where $\sum_s p_s = 1$.  The choice of $\hat{U}$, $ \mathcal{K}$ could of course vary from layer to layer.  An example of such a circuit is depicted in \cref{fig:cartoon}. Note that Pauli noise is particularly relevant to near term experiments, as even when the physical hardware has a more complicated error model, experiments often employ Pauli twirling (or randomized compiling) so that all of the errors can be modeled as a Pauli channel \cite{Wallman2016}.

At the outset, the combination of matchgate circuits and Pauli noise is no longer easily simulable.  While each ingreident can be efficiently simulated on their own, they rely on different techniques that would seem to be incompatible.  As noted, the unitary dynamics can be mapped onto free fermions because the dynamics preserve the Gaussianity of the fermionic state. However, the noise operators do not map to simple operators under JW, and moreover the noise does not preserve Gaussianity of the density matrix. Alternatively, Pauli noise preserves Cliffords, and therefore can be efficiently simulated as a Clifford circuit. However, matchgate circuits can inject an arbitrary amount of non-stabilizerness (or magic) \footnote{This is because the T-gate $\hat T = \exp(i \pi \hat \sigma^z/8)$ is in fact a matchgate, and so any amount of magic can be present in a matchgate circuit \cite{Bravyi2005}. We stress that this does not mean that matchgates are universal, as they do not include the entire Clifford group. There is a different notion of ``fermionic magic'' which requires a pure non-Gaussian resource state to make matchgates universal \cite{Hebenstreit2019}.}, and so this immediately breaks simulability of the dynamics as a Clifford circuit.

Progress was made on this problem in Ref.~\cite{Pocklington2025}, where it was shown that, while the state does not remain Gaussian under the noisy evolution, it can be written as a convex sum of Gaussian states \cite{Melo2013,gonzalez2025}.  As a result, dynamics can be efficiently described using a stochastic equation of motion for the fermionic covariance matrix. 

Here, we make further progress by noting that, when the noise is Pauli, the above stochastic equation of motion can be {\it exactly} re-averaged. If we define Majorana modes $\hat \gamma_{2m} = \hat c_m + \hat c_m^\dagger, \hat \gamma_{2m + 1} = i(\hat c_m - \hat c_m^\dagger)$ and a covariance matrix $\Gamma_{mn} = \langle \hat \gamma_m \hat \gamma_n \rangle$, then we can show that the evolution of the covariance matrix under the noise channel $\mathcal{K}$ is given by
\begin{align}
    (\mathcal{K}_\mathrm{cov}[\Gamma])_{mn} &= f_{mn}(\vec p) \Gamma_{mn}. \label{eqn:cov_update}
\end{align}
We have denoted $\mathcal{K}_\mathrm{cov}: \mathbb{C}^{2N \times 2N} \to \mathbb{C}^{2N \times 2N}$ as the induced map between covariance matrices derived from the quantum channel $\mathcal{K}$. The constants $f_{mn} \in \R$ can be calculated exactly for any (Pauli) noise model. 
We derive this expression for arbitrary Pauli noise in \cref{app:A}, and also show explicitly how to calculate the constants $f_{mn}$. 

The simplicity of \cref{eqn:cov_update} reveals two surprises.  The first is that the covariance matrix updates close on themselves, despite the fact that the state does not remain Gaussian. This is a new entry to the set of fermionic master equations with closed hierarchies of moments \cite{Shtanko2021}, complementing previously known results for linear jump operators \cite{Prosen2008,Prosen2010} and quadratic, Hermitian jump operators \cite{Znidaric2010,Eisler2011,Horstman2013}. The second surprise is that not only do the moments close on themselves, but in the Majorana basis $\mathcal{K}_\mathrm{cov}$ is diagonal: following an arbitrary Pauli noise channel, the two-point function $\langle \hat \gamma_m \hat \gamma_n \rangle$ depends only on its previous value and the noise strength, and not on any other correlation function. In general, this holds for all higher moments written in the Majorana basis.

To give a concrete example, the primary noise model we study is given by equal strength spatially uniform dephasing in the X and Y directions, where 
\begin{align}
    \mathcal{K}[\hat \rho] &= \prod_i \mathcal{K}_{i,x} [\mathcal{K}_{i,y} [ \hat \rho ]], \label{eqn:xy_channel} \\
    \mathcal{K}_{i,\alpha}[\hat \rho] &= p \hat \sigma_i^\alpha \hat \rho \hat \sigma_i^\alpha + (1-p) \hat \rho  .
\end{align}
For this particular channel, the constants $f$ are given by (see \cref{app:A})
\begin{align}
    f_{mn}(p) &= \left\{
    \begin{array}{cc}
      (1 - 2 p)^{2|\floor{m/2}-\floor{n/2}|}   & \floor{m/2} \neq \floor{n/2}  \\
       (1 - 2 p)^{2(1 - \delta_{mn})}  & \floor{m/2}=\floor{n/2}
    \end{array}
    \right. .
\end{align}
This result allows us to calculate arbitrary observables exactly for all times.

Moreover, we show that the result can be generalized to the continuous time version of the problem with a Lindblad generator as opposed to a finite noise channel. Given any Hamiltonian of the form of \cref{eqn:Ham} and Pauli noise, we are also capable of simulating a time continuous process described by a Lindblad master equation \cite{Lindblad1976,Gorini1976}:
\begin{align}
    \partial_t \hat \rho &= -i[\hat H, \hat \rho] + \sum_{s \in \{0,1\}^{2n}} \gamma_{s} \D[\hat K_s] \hat \rho,
\end{align}
where $\D[\hat z] \hat \rho \equiv \hat z \hat \rho \hat z^\dagger - \frac{1}{2} \{ \hat z^\dagger \hat z, \hat \rho \}$, and we have reused the definition of $\hat K_s$ as given in \cref{eqn:Kraus_op}. 

\section{Transverse Field Ising Model in a Noisy Magnetic Field}
\label{sec:IsingModel}
The model we wish to study here is a transverse field Ising model subject to spatially uniform noise. We consider a Hamiltonian with an Ising interaction in the $X$ direction, and a transverse field in the $Z$ direction:
\begin{align}
    \hat H &= \sum_i -J \hat \sigma_i^x \hat \sigma_{i + 1}^x +  g \hat \sigma_i^z . \label{eqn:TFIM}
\end{align}
We choose to study this system because it can be exactly solved using the techniques outlined above in \cref{sec:solution}, as well as admits a phase transition at $|g/J| = 1$, allowing us to study how critical states behave in the presence of noise exactly. Unless otherwise noted, we will work in the thermodynamic limit, in which case the Hamiltonian can be diagonalized in terms of plane wave momentum modes (see \cref{app:B}):
\begin{align}
    \hat H &= \sum_{k} \epsilon_k \hat \beta_k^\dagger \hat \beta_k , \\
    \hat \beta_k &= \cos(\theta_k/2) \hat d_k + i \sin(\theta_k/2) \hat d_{-k}^\dagger ,
\end{align}
where $\tan \theta_k = \frac{J\sin(k)}{g - J \cos(k)}$ and $\epsilon_k = 2 \sqrt{J^2 + g^2 - 2 g J \cos(k) }$. Here, $\hat d_k$ is a fermion momentum mode with momentum $k$: $\hat d_k = \frac{1}{\sqrt{N}} \sum_j e^{i jk} \hat c_j$.

When $|g/J| = 1$, the spectrum becomes gapless for the long wavelength $k = 0$ mode. Moreover, the ground state at this point has long-range order, characterized by an algebraically decaying spatial correlation function. In the Majorana basis (for $m \leq n$)
\begin{align}
    \langle \hat \gamma_{2m} \hat \gamma_{2n + 1} \rangle &= \frac{2i}{\pi} \frac{1}{1 - 2(m-n)}, \\
     \langle \hat \gamma_{2m} \hat \gamma_{2n} \rangle &=  \langle \hat \gamma_{2m + 1} \hat \gamma_{2n + 1} \rangle = 0.
\end{align}

We want to understand the fate of the critical properties of this model in the presence of noise. Motivated by the previous literature on decohered and measurement altered critical states \cite{Garratt2023,Murciano2023,Zou2023}, we consider the case where one first prepares the critical ground state, and then applies only the noise channel with no coherent (Hamiltonian) dynamics. We couple each spin to a local Markovian reservoir described by an equal probability of flipping the spin from up to down or down to up. This is given by the Lindblad equation
\begin{align}
    \mathcal{L}_{XY} &= \sum_{i} \D[\hat \sigma_i^-] + \D[\hat \sigma_i^+]  = \frac{1}{2} \sum_{i}   \D[\hat \sigma_i^x]  + \D[\hat \sigma_i^y] . \label{eqn:XYnoise} \\
    \partial_t \hat \rho &= -i \theta [\hat H, \hat \rho] + \gamma \mathcal{L}_{XY} \hat \rho. \label{eqn:XY_ME}
\end{align}
 We define the binary variable $\theta \in \{0,1\}$; unless otherwise specified, we consider the case where $\theta = 0$ so there is noise-only dynamics (however, we consider $\theta = 1$ in \cref{sec:WithHam}). \cref{eqn:XYnoise} describes a set of independent, infinite temperature, infinite bandwidth reservoirs. Equivalently, this describes driving each spin with a classically stochastic magnetic field in the $X$ and $Y$ directions that is identical and independent on each lattice site. Evolution under the master equation \cref{eqn:XY_ME} (with $\theta = 0$) for a time $t$ is identical to applying the noise channel defined in \cref{eqn:xy_channel} with $p = (1 - e^{-\gamma t/2})/2$. We are not in principle limited to considering spatially or temporally uniform dephasing, nor did the strength of the dephasing in the $X$ and $Y$ directions need be equivalent, but it was chosen for simplicity and by experimental motivation.

Despite the fact that this is an interacting many body spin model, we are able to get \textit{exact, closed form} expressions for the dynamics of \textit{all} correlation function of the fermionic operators, see \cref{app:B} for more details. The equation of motion for the two-point correlation functions is given by:
\begin{align}
    \partial_t \langle \hat \gamma_{2m} \hat \gamma_{2n + 1} \rangle = -\gamma (|m-n| + \delta_{mn}) \langle \hat \gamma_{2m } \hat \gamma_{2n + 1} \rangle . \label{eqn:noiseStrength} \\
    \implies \langle \hat \gamma_{2m} \hat \gamma_{2n + 1} \rangle (t) = \frac{2i}{\pi} \frac{ e^{-(\gamma|m-n| + \delta_{mn})t}}{1 - 2(m-n)} . \label{eqn:2ptEOM}
\end{align}
This takes a surprisingly simple form; however, we again stress that even though the equations of motion for the two point functions close on themselves, the state will generally evolve to something \textit{highly} non-Gaussian. Using Wick's theorem to try to calculate higher point correlation functions could result in values that diverge exponentially from their true values, see \cref{app:A.3}.

\section{Decohered Critical Phenomena}
\label{sec:CritPhenom}

In this section, we will try to understand the effect of just noise ($\theta = 0$) on an initial critical state.

\subsection{Emergent length scale}

Using the exact solution for the equation of motion of the fermionic degrees of freedom, we can now ask what physical understanding we can get from this model. One of the most striking things that emerges immediately from the form of \cref{eqn:2ptEOM} is that the fermions have an emergent length scale for any time $t > 0$, given by
\begin{align}
    \xi_F &= \frac{1}{\gamma t}
\end{align}
The reason that this is surprising is that the initial state was scale invariant, and the noise channel we applied is local and finite depth, which cannot create a length scale from a scale invariant state 
\footnote{This is because Pauli noise can be described by a unitary interaction with an ancilla qubit that is then traced out. Because this is a finite depth channel, it is impossible to alter any features of the long-range correlations characteristic of critical states. } .
This apparent tension is resolved by noting that the actual spins never develop a length scale. For example, the correlations in the Ising direction obey
\begin{align}
    \langle \hat \sigma_m^x \hat \sigma_n^x \rangle \sim |m-n|^{-1/4} e^{- \gamma t} . 
\end{align}
This provides an interesting juxtaposition: on the one hand the physical spins are scale invariant, but on the other hand the degrees of freedom which diagonalize the Hamiltonian do develop a length scale which is tending to zero as a function of time. 

Given that any real experiment will measure the spins as opposed to the fermionic operators, a natural question is whether this emergent length is ever observable or experimentally relevant. As we will show, it is in fact observable with an extremely simple experimental apparatus by probing a kind of ``effective temperature''.

\begin{figure}[t!]
    \centering
    \includegraphics[width = \linewidth]{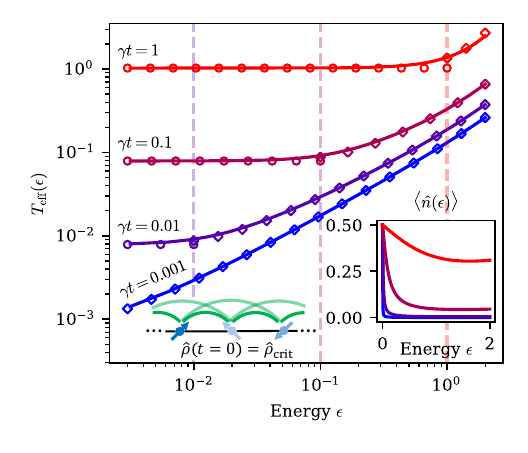}
    \caption{Effective temperature $T_{\rm eff}(\epsilon)$ of the decohered critical Ising ground state. 
    For all curves, the initial condition is the ground state of the Hamiltonian in \cref{eqn:TFIM} at $g = J = 1$.  One then applies the decoherence channel (\cref{eqn:XYnoise,eqn:XY_ME} with $\theta = 0$); 
    different curves correspond to different times $\gamma t$ as indicated. For longer evolution times, one sees clearly the emergence of a constant effective temperature at low energies. Solid lines are the exact $T_{\rm eff}(\epsilon)$ (c.f.~\cref{eqn:Teff}). Open circles (diamonds) are the approximation given in \cref{eqn:TeffLimits} for the long (short) times. Vertical dashed lines denote the energy where the approximation switches its functional form.   Inset: quasiparticle occupation corresponding to the $T_{\rm eff}$ in the main plot, using \cref{eqn:NkFull}.}
    \label{fig:effectiveTemp}
\end{figure}

\subsection{Effective temperature}

As noted above, the length scale is emergent only when looking at the fermionic degrees of freedom and not the physical spins. While the spins may be the natural microscopic degree of freedom, the energetics are naturally described by the fermions since they describe the quasiparticle excitations above the ground state. Despite the fact that there is no Hamiltonian dynamics, knowledge of the Hamiltonian is encoded in the initial state which has knowledge of the quasiparticle excitations. This motivates looking at an observable with energy resolution to try and see this length scale. The simplest such observable is the expectation value of the quasiparticle number at a given energy/wavevector.

We can define the set of number operators $n_k = \langle \hat \beta_k^\dagger \hat \beta_k \rangle$ and ask how they evolve in time under the applied noise (c.f.~\cref{eqn:XYnoise,eqn:XY_ME} with $\theta = 0$). Now, before performing the explicit calculation, there is reason to believe that the structure of the noise coupled with a critical state can lead to extremely counter-intuitive effects. For a real model in thermal equilibrium at some temperature $T$, low energy modes will be more populated than high energy modes. In our setting the noise is perfectly Markovian, and so one might assume it does not care at all about the energy of the modes it is exciting. However, because of the string operators, the noise is sensitive to long-range correlations: the longer distance the correlation function, the more strongly the noise couples [c.f.~\cref{eqn:noiseStrength}]. This implies the noise is sensitive to distance, or equivalently momentum, which in turn gives a path to energy selectivity while still being infinitely broadband. Moreover, because the long-range order of the critical state stems from the spatial structure of the low-energy modes, this gives a mechanism by which the low-energy modes are preferentially excited by a broadband source, mimicking a true temperature.

Before considering the full time dynamics, it is informative to consider both the limit $|k|/(\gamma t) \ll 1$ and $|k|/(\gamma t) \gg 1$. Equivalently, this is the same as asking if the fermionic length scale $\xi_F  = (\gamma t)^{-1}$ is longer or shorter than a given wavelength $\propto |k|^{-1}$. For short times, the length scale is still longer than the wavelength, and we find that (see~\cref{app:B.1}):
\begin{align}
    n_k(t) &= \frac{2}{\pi} \left(  \frac{J}{\epsilon_k} + \frac{\epsilon_k}{8 J}\right) \gamma t + \mathcal{O}(\gamma t/|k|)^3 . \label{eqn:approx1}
\end{align}
Since we are considering a long-wavelength expansion $k \ll 1$, then $\epsilon_k/J \ll 1$ and the rate at which quasiparticles are being injected by the noise scales inversely with their energy, in line with our intuition that the noise couples most strongly to the low energy modes. In the opposite limit of long time where the length scale is much shorter than the wavelength, we have that
\begin{align}
    n_k(t) &= \frac{1}{2} + \frac{\epsilon_k}{4 \pi J}   (1 - e^{-\gamma t} - \coth(\gamma t/2))  + \mathcal{O}\left(\frac{|k|}{\gamma t} \right)^3 . \label{eqn:approx2}
\end{align}
Note that for long times, $(1 - e^{-\gamma t} - \coth(\gamma t/2)) = -3 e^{-\gamma t} + \mathcal{O}(e^{- \gamma t})^2$. This implies that at long times, we have exponential relaxation at a rate $\gamma$ which is the same for all modes, but they behave as if they are relaxing from an energy dependent initial condition of $1/2 - 3\epsilon_k/(4 \pi J)$. 

Both the long and short time dynamics can be understood heuristically by thinking about the interplay between the noise and the fermionic length scale $\xi_F$. When the length scale $\xi_F \gg k^{-1}$ the wavelength, then the noise couples extremely strongly to this wavevector, and so the population grows faster than one would expect, like $1/\epsilon_k$.  Very heuristically, we can think of this as a two-step relaxation process. First, at very short times there is some initial heating that depends strongly on the nature of the long range order and is very sensitive to the energy of a given quasiparticle mode. This process is cutoff when the length scale is less than the wavelength of the quasiparticle, $\xi_F \lesssim |k|^{-1}$. After this, there is uniform relaxation to infinite temperature at a rate independent of the energy, but with the initial fast dynamics leaving an imprinted ``initial condition'' of the exponential relaxation. It is this initial imprint (which scales roughly linearly with energy) that gives rise to the effective temperature. 

\begin{figure}[t]
    \centering
    \includegraphics{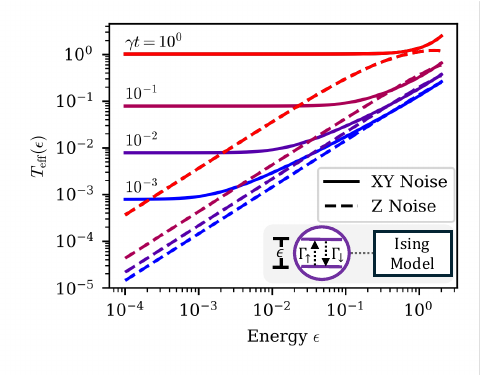}
    \caption{Effective temperature as shown in \cref{fig:effectiveTemp} of the decohered ground state of the critical Ising model with $g=J$ for variable decoherence time $\gamma t \in \{0.001,0.01,0.1,1 \}$. Solid lines denote decoherence in the $X$ and $Y$ directions [c.f. \cref{eqn:XYnoise}] identical to the data in \cref{fig:effectiveTemp}. Dashed lines show decoherence for the same time in the $Z$ direction [c.f. \cref{eqn:Znoise}]. Inset: experimental technique to measure the temperature using a type of noise spectroscopy. The probe qubit (purple) with a splitting $\Delta$ is weakly coupled to the Ising chain, which induces incoherent transitions at a rate $\Gamma_{\uparrow/\downarrow}$.}
    \label{fig:teff_v_noise}
\end{figure}

The full closed form solution for the dynamics at all times is given  by (see \cref{app:B.1}):
\begin{widetext}
\begin{align}
    n_k(t) &= -\frac{i}{\pi}\cosh(t/2)  \left[ \tanh^{-1}\left( e^{(-i |k| - \gamma t)/2} \right) - \tanh^{-1}\left( e^{(i |k| - \gamma t)/2} \right) \right]  + \frac{1}{\pi} |\sin(k/2)|(1 - e^{- \gamma t}) + \frac{1}{2}. \label{eqn:NkFull}
\end{align}
\end{widetext}
Given a value of the quasiparticle population, we can use this to define a kind of effective temperature that depends on both time and energy by asking, given a quasiparticle density, what temperature would this correspond to in an equilibrium system. This temperature is defined by
\begin{align}
    T_{\mathrm{eff}}(k,t) &= \frac{\epsilon_k}{\log(1 - n_k) - \log(n_k)}.  \label{eqn:Teff}
\end{align}
We can use the previously derived approximations for the limiting behavior of $n_k$ [c.f.~\cref{eqn:approx1,eqn:approx2}] to find the limiting behavior of the effective temperature (see \cref{fig:effectiveTemp}):
\begin{align}
    T_{\mathrm{eff}}(k,t) &\sim \left\{ \begin{array}{cc}
       \pi J \frac{e^{2 \gamma t} - e^{\gamma t}}{3e^{\gamma t} - 1}  & \gamma t \gtrsim |k| \\
        -4\epsilon_k \log^{-1} \left[ \left( \frac{\epsilon_k}{2J} + \frac{J}{\epsilon_k} \right) \frac{\gamma t}{\pi}  \right] & \gamma t \lesssim |k|    \end{array} \right.  . \label{eqn:TeffLimits}
\end{align}
Note that for long times, when the length scale $\xi_F < |k|^{-1}$, then the effective temperature becomes effectively independent of the energy/momentum and the system looks as if it has reached a thermal equilibrium. In this way, one can directly observe the length scale by measuring the ``temperature'' as a function of momentum and seeing at what wavevector the system begins to look nonequilibrium. A sketch of how this experiment might work is given in \cref{fig:cartoon} as well as the inset of \cref{fig:teff_v_noise}. To probe the particle density as a function of energy, one can attach a probe qubit, denoted by operators $\hat \sigma_P^{x,y,z}$ to the edge of the Ising model with a weak XX coupling:
\begin{align}
    \hat H &= \frac{\Delta}{2} \hat \sigma_z^P + \hat H_{\mathrm{Ising}} + \lambda(\hat \sigma_P^+ \hat \sigma_1^- + \hc) ,
\end{align}
where $H_{\mathrm{Ising}} $ is the Hamiltonian in \cref{eqn:TFIM}. In the limit that the coupling is extremely weak $\lambda \ll J$, then only quasiparticles with an energy exactly at the probe qubit splitting $\Delta$ can tunnel in, and so the effective dynamics can be described by rates $\Gamma_{\uparrow,\downarrow}$ for the probe qubit to flip from down to up or up to down, respectively, which depend only on the coupling strength $\lambda$ and the quasiparticle density at the energy $\Delta$. In the long time limit, the probe qubit will equilibrate so that the number of quasiparticles is encoded in the expectation value of $\hat \sigma _P^z$ (see \cref{app:D.1})
\begin{align}
    \langle \hat \sigma_P^z \rangle(t \to \infty) &= (2n_k - 1), \ \ \epsilon_k = \Delta .
\end{align}
Thus, if one simply performed thermometry on the probe qubit and asked its ``temperature'' given the splitting frequency $\Delta$ and the population imbalance, one would exactly recover $T_\mathrm{eff}(\Delta)$ of the Ising model, as desired. More details can be found in \cref{app:D}. 

This protocol is reminiscent of schemes to cool many-body systems by extracting energy into an ancilla at a fixed frequency \cite{Metcalf2020,Polla2021,Mi2024,Matthies2024,Kishony2025}; the difference here is that we work in a regime where instead of attempting to actually extract a large amount of energy, we are in a regime where the actual energy transfer is small enough to be negligible to the original system, as we aren't trying to cool the system and instead are trying to probe the state.

\begin{figure}[t]
    \centering
    \includegraphics{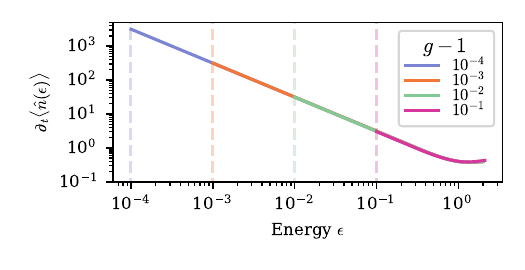}
    \caption{    
    Instantaneous rate of change of the quasiparticle number at $t = 0$ as a function of energy under the decoherence channel \cref{eqn:XYnoise,eqn:XY_ME} with $\theta = 0$. The initial condition is the ground state of the Ising model \cref{eqn:TFIM} for the given value of $g$ with $J = 1$. Vertical dashed lines denote the corresponding energy gap.}
    \label{fig:totalNDeriv}
\end{figure}

{\subsection{Divergent quasiparticle production}}
\label{subsec:divergentQP}
The emergence of a temperature comes after applying the noise channel for a relatively long time, so that the fermionic length scale $\xi_F$ becomes comparable to a handful of lattice sites. We instead now consider what phenomena can be seen in the extremely short time limit. The most extreme short time data is the derivative $\partial_t n_k$ in the ground state. The reason to believe that this quantity might prove interesting is that, when the state is critical, we know the noise causes a sharp jump in the QP number. In fact, one can show that to leading order in $\epsilon_k$ (see \cref{app:B.4})
\begin{align}
    \partial_t n_k|_{t = 0} &= \frac{2 \gamma}{\pi \epsilon_k} + \mathcal{O}(\epsilon_k^0) .
\end{align}
This scaling is shown in \cref{fig:totalNDeriv}, where the inverse scaling of the derivative with the energy is immediately apparent. Such a divergence in the derivative at short times manifests in a total particle number that, starting from the critical state, behaves as (see \cref{app:B.4}):
\begin{align}
    n &= \int \frac{\dd k}{2 \pi} n_k \sim \left\{
    \begin{array}{cc}
       -\frac{1}{\pi^2}\gamma t \log \gamma t  & \gamma t \ll 1 \\
       \frac{1}{2} - \frac{38}{9 \pi^2} e^{-\gamma t}  & \gamma t \gg 1
    \end{array}
    \right. \label{eqn:ntot_TFIM} .
\end{align}
This divergence is in keeping with the notion that critical states are very fragile to noise: defects are created super-linearly at short times. At long times, it also exactly mimics the notion that there is a quick ``jump'' followed by uniform relaxation. 

Because the total quasiparticle number grows superlinearly at short times, this reflects another experimental signature of coupling the noise to a critical state. Moreover, it is again experimentally measurable. Instead of weakly coupling a probe, an even simpler protocol is imagined: after applying the noise, one turns the Ising Hamiltonian back on, and performs an adiabatic ramp from the critical point ($g/J = 1$) to either the perfect ferromagnet ($g/J = 0$) or the perfect paramagnet ($g/J=\infty$). If one ramps to the paramagnet, then projectively measuring $Z$ on every lattice site and counting how many spins are in the $|\uparrow\rangle$ state gives the number of QPs (see \cref{app:D}). Alternatively, ramping to the ferromagnet, projectively measuring in the $X$ basis, and counting domain walls also gives the number of QPs. 

\begin{figure}[t]
    \centering
    \includegraphics{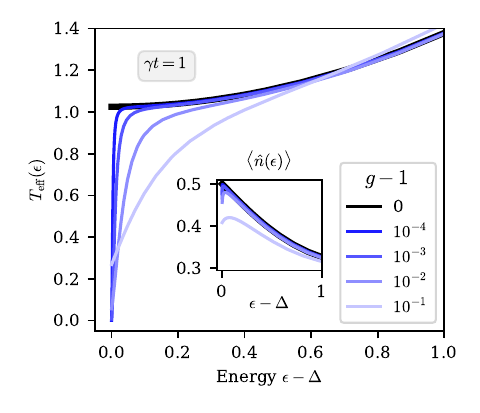}
    \caption{Effective temperature of the decohered ground state away from the critical point. $\Delta = 2|g-1|$ corresponds to the energy gap, and $\epsilon - \Delta$ the energy above the gap. The initial condition is the ground state of the Ising model \cref{eqn:TFIM} for the given value of $g$ with $J = 1$. The state is given by time evolution under \cref{eqn:XYnoise,eqn:XY_ME} with $\theta = 0$ for a time $\gamma t = 1$. Solid black lines ($g-1 = 0$) denotes the critical point as depicted in \cref{fig:effectiveTemp} for comparison. Inset: quasiparticle densities corresponding to the effective temperature in the main plot.}
    \label{fig:temp_v_g}
\end{figure}

\bigskip
{\section{Dynamics away from the critical point}}
\label{sec:NonCrit}

In this section, we will compare the decohered critical state to decoherence applied to a gapped ground state. We will still only consider evolution under the noise only dynamics \cref{eqn:XYnoise,eqn:XY_ME} with $\theta = 0$.

\bigskip
{\subsection{Lack of emergent temperature}}

We have seen that applying the infinite temperature noise channel to the critical ground state induces an effective temperature up to a length scale $\xi_F$. We now ask how important it was that the initial condition was a critical state having long range order. Intuitively, one might expect that the dynamics will be quite different if the initial state already has a length scale. The sudden ``jump'' in quasiparticle number at small wavevector and short times was directly due to the fact that there was long range order and the noise could couple extremely strongly to these low frequency modes. When the system is gapped, this is no longer the case.

This intuition is supported by the exact calculations for the Ising model, which can be seen in \cref{fig:temp_v_g}, where the effective temperature is plotted for varying value of $g/J - 1$ after decohering for a time $\gamma t = 1$. We note that, even after moving only slightly away from the critical point to $g = 1.1J$, the lightest curve in the plot, the effective temperature for low energies looks extremely different than that of the critical ground state. Instead of being flat as a function of energy, it is now growing linearly for small energies, making it highly nonequilibrium. 

\bigskip
{\subsection{Logarithmically diverging quasiparticle production}}

Recall from \cref{subsec:divergentQP} that when initializing the quantum state exactly at the critical point, then the noise channel injects QPs at a rate that is diverging at short times. Intuitively, this is a result of the fact that exactly at $t=0$, the quantum state has algebraically decaying correlation functions. Since the noise strength is growing linearly with separation, these two effects ultimately cause a divergence. Alternatively, when the initial condition is no longer critical, the correlations decay exponentially and therefore we expect the rate at which quasiparticles are being injected should be bounded. We wish to understand the leading order behavior of $\partial_t n\vert_{t = 0}$ as a function of $g-1$, working in units where $J = 1$. When $g = 1$ we know this value must diverge, but it is informative to understand how this works. 

This value can be represented by the following momentum space integral (see \cref{app:B}):
\begin{align}
    \partial_t n \vert_{t = 0} &\sim \int_{-\pi}^{\pi} \frac{\dd k}{2 \pi} \int_{-\pi}^{\pi} \frac{\dd k'}{2 \pi} \frac{\sin^2([\theta_k - \theta_k']/2)}{(k-k')^2}.
\end{align}
which is valid for all values of $g-1$. As expected, this quantity can only diverge directly at the critical point, as the integrand is strictly bounded above by $(1-g)^{-2}/4$. Utilizing the exact solution, one can show that to leading order
\begin{align}
    \partial_t n \vert_{t = 0} &= -\frac{\gamma}{\pi^2} \log(g-1) + \mathcal{O}(g-1)^0.
\end{align}
Therefore, the rate at which QP are being injected into the system is diverging logarthmically as one approaches the critical point. This is incredibly intuitive when looking at \cref{fig:totalNDeriv}, where one can see that the energy resolved QP number is being injected at a rate $\sim 1/\epsilon$. Integrating this over all energy, recalling the flat density of states at low energy and that the gap is set by $g-1$ gives a logarthmic divergence.

Together, the lack of temperature and lack of divergence highlights the fact that there is an extremely nontrivial interplay between criticality and noise. Decohered critical states effectively thermalize under local Markovian noise in a way that gapped ground states do not. We believe this is a result of the nature of how noise can couple to scale invariant states that is distinctly different from how it couples to states with exponentially decaying correlations. Moreover, we posit that this result may be more general than just the transverse-field Ising model, and instead reflect characteristics of more general decohered critical models. Given generic local noise operators, if the low energy excitations are described by long-wavelength fluctuations then the noise will be able to degrade these correlations much more quickly than short range correlations. This heuristic does not depend on the microsopic details of the Ising model, and it would be of interest to see if this holds more generally (which we explore briefly in \cref{sec:critXX}). Additionally, it could point to an interesting experimental signature of criticality.

\bigskip
\section{Noise With Coherent Dynamics}
\label{sec:WithHam}

In this section, we will compare the results derived in \cref{sec:CritPhenom} to the case where there are also coherent Hamiltonian dynamics. This can be done either in a time continuous process or as a discrete circuit of interleaved coherent dynamics with discrete noise channels as depicted in \cref{fig:cartoon}. We will consider the time-continuous version here for simplicity and take dynamics given by \cref{eqn:XYnoise,eqn:XY_ME} with $\theta = 1$. 

In the case that there are both coherent dynamics and noise, we no longer have closed form expressions for the time traces of the quasiparticle number; however, there are still closed expressions for the equations of motion of the fermionic two point functions, and we can access very large system sizes using numerics. In order to understand how these curves converge in the thermodynamic limit, it is important to attempt to mitigate boundary effects that should be asymptotically irrelevant to the bulk physics.

In order to do this, the boundary conditions we impose will be relevant. In fact, in \cref{app:A} we show that while the low energy physics with periodic boundary conditions (PBC) converge as $N^{-1}$; when there are open boundary conditions (OBC) it only converges as $N^{-1/2}$. Therefore, it is often preferable to work directly with PBC. There is some subtlety here, though, as the PBC Hamiltonian is interacting:
\begin{align}
    \hat H_{\mathrm{PBC}} &= -J \hat \sigma_1^x \hat \sigma_N^x - J\sum_{i = 1}^{N-1} \hat \sigma_i^x \hat \sigma_{i + 1}^x +  g \sum_{i = 1}^N \hat \sigma_i^z \nonumber \\
    &= J (\hat c_N^\dagger - \hat c_N)(\hat c_1 + \hat c_1^\dagger) \hat \Pi \nonumber \\
    & \ \ - J\sum_{i = 1}^{N-1} (\hat c_i^\dagger - \hat c_i)(\hat c_{i + 1} + \hat c_{i  +1}^\dagger) + 2 g \sum_{i = 1}^N \hat c_i^\dagger \hat c_i , \label{eqn:PBC}
\end{align}
where $\hat \Pi = \prod_{i = 1}^N (-1)^{\hat c_i^\dagger \hat c_i} = \prod_{i = 1}^N \hat \sigma_i^z$ is the total parity operator. This is also the $\Z_2$ symmetry of the Ising Hamiltonian and is a weak symmetry of the noisy dynamics \cite{Buca2012}. Now, the Hamiltonian is formally interacting, but the interaction comes from a conserved quantity $[\hat H_{\mathrm{PBC}}, \hat \Pi] = 0$. Therefore, we can replace the operator $\hat \Pi$ with its expectation value and diagonalize the Hamiltonian in the different parity sectors. When there are an even number of quasiparticles, then $\hat \Pi = 1$ and so $\hat H_{\mathrm{PBC}}$ is a free-fermion Hamiltonian with anti-periodic boundary conditions (APC); i.e., one can imagine threading a $\phi = \pi$ flux through the loop. On the other hand, when there are an odd number of quasiparticles, then $\hat \Pi = -1$, and so $\hat H_{\mathrm{PBC}}$ is again free and has PBC. 

Therefore, the covariance matrix dynamics are still simple and can be solved exactly in each parity sector. The next step is to realize that, while the noise can change parity, it can never generate coherences between different parity sectors. This is in keeping with fermion super-selection rules, and formally is a result of the fact that parity is a weak symmetry of the dynamics. Therefore, we can observe that, not only can the state be written as a convex sum of Gaussians, it can also be written as a convex sum of each fixed parity sector. Specifically, if we consider a pure state $|\Gamma \rangle$ as a Gaussian free-fermion state with covariance matrix $\Gamma$, then its parity is given by $\Pf(\Gamma)$ where $\Pf(\bullet)$ stands for the Pfaffian. We can write the total density matrix at all times in the form:
\begin{align}
    \hat \rho(t) &= \hat \rho_+(t) + \hat \rho_-(t) , \\
    \hat \rho_{\pm}(t) &= \sum_{\Gamma| \Pf(\Gamma) = \pm 1} p_{\pm}(\Gamma, t) |\Gamma \rangle \langle \Gamma | .  
\end{align}
Here, $\hat \rho_\pm$ are (unnormalized) fixed parity convex sums of Gaussians with probability distributions over the space of covariance matrices given by $p_\pm$. We can use these to define even and odd covariance matrices $\Gamma_\pm$ given by 
\begin{align}
    (\Gamma_{\pm})_{mn} &= \tr ( \hat \rho_{\pm} \hat \gamma_m \hat \gamma_n ) .
\end{align}
Now, to understand what the noise does, we can observe that it will mix the even and odd parity sectors into each other. Specifically, if there is a finite depth channel with local probability $p$ to perform a parity-reversing jump, then we can observe that the total channel can be written as
\begin{align}
    \mathcal{K}[\hat \rho] &=  (1-p)\mathcal{K}_+[\hat \rho] + p\mathcal{K}_-[\hat \rho] ,
\end{align}
where $\mathcal{K}_\pm[\bullet] = \sum_{i} \hat K_{i,\pm} \bullet \hat K_{i,\pm}^\dagger$ with $[\hat K_{i,+}, \hat \Pi] = 0$ commuting with parity, and $\{ \hat K_{i,-}, \Pi \} = 0$ anticommuting. Given the local noise model we have been considering of equal probability $X$ and $Y$ dephasing, then $\hat K_{i,+}$ are proportional to all strings of Paulis of even length, and vice versa for $\hat K_{i,-}$ and odd length strings because all jumps flip parity. Because the jump probabilities after separating into even and odd sectors remain independent of the state, it is still possible to exactly re-sum the stochastic equations of motion generated under these trajectories, which is explained in more detail in \cref{app:A}. Moreover, we can show that the overall effect is that the coupling rates between the two parity sectors is weakly dependent on system size because the probability of having an even or odd length string depends weakly on the total number of lattice sites. Overall, if previously the covariance matrix $\Gamma_{mn}$ obeyed the equation of motion $\partial_t \Gamma_{mn} = -\gamma f_{mn} \Gamma_{mn}$ where $\gamma$ is a rate. Once we consider separately the different parity sectors we find that this gives the channel:
\begin{widetext}
\begin{align}
    \left( 
    \begin{array}{c}
        (\Gamma_+)_{mn} \\
        (\Gamma_-)_{mn}
    \end{array}
    \right)(t) &= 
    e^{- \gamma N t} \left( 
    \begin{array}{cc}
       \cosh(\gamma t(N - f_{mn}) & \sinh(\gamma t(N - f_{mn}) \\
       \sinh(\gamma t(N - f_{mn}) & \cosh(\gamma t(N - f_{mn})
    \end{array}
    \right)
    \left( 
    \begin{array}{c}
        (\Gamma_+)_{mn} \\
        (\Gamma_-)_{mn}
    \end{array}
    \right)(0) .
\end{align}
\end{widetext}
We can check that if there are no other dynamics, then $[ (\Gamma_+)_{mn} + (\Gamma_-)_{mn} ] (t) = e^{-\gamma f_{mn} t}[ (\Gamma_+)_{mn} + (\Gamma_-)_{mn} ] (0)$, recovering the previous results. However, this will not hold exactly when $\Gamma_\pm$ evolve under different Hamiltonian dynamics.

\begin{figure}[t]
    \centering
    \includegraphics{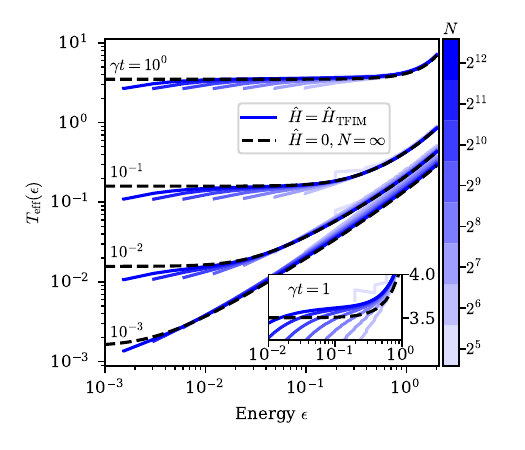}
    \caption{Interplay between coherent and incoherent dynamics, where the initial condition is the critical ground state. The blue curves show time evolution under the master equation \cref{eqn:XYnoise,eqn:XY_ME} with $\theta = 1$ for system sizes varying from $N=32$ (lightest) to $N = 4096$ (darkest), with periodic boundary conditions. The black dashed curves show time evolution under the master equation \cref{eqn:XY_ME} with $\theta = 0$ in the thermodynamic limit $N = \infty$ [c.f.~\cref{fig:effectiveTemp}]. Inset: zoomed in details for $\gamma t = 1$.}
    \label{fig:Hamiltonian Data}
\end{figure}

Now that we understand how to solve periodic boundary conditions, we can consider what happens under the combined coherent and noisy channels. This is shown in \cref{fig:Hamiltonian Data}, for system size varying from $N = 32$ to $N = 4096$. Each blue curve is initialized at the critical ground state, and then evolves under the critical Ising Hamiltonian and noise. The black dashed curves are the closed form results in \cref{eqn:NkFull} when there is only noise. One can observe that the curves with the coherent Hamiltonian dynamics are asymptotically converging to something qualitatively similar to the noise only dynamics, but there are some small quantitative differences, as can be seen in the inset. One can also see that the low energy physics is converging extremely slowly in system size, even though the high energy QP numbers converge rapidly.

\section{Other Noise Types}
\label{sec:otherNoise}

Here, we will briefly consider alternative decoherence channels, and see the effect they have on the phenomena we have observed both at and away from the critical point. We will again only consider pure noise dynamics where $\theta = 0$.

\subsection{Transverse field noise}

First, we consider the more commonly studied noise model corresponding to decoherence in the $Z$ direction \cite{Patane2008,Nalbach2015,Nigro2019}. An equivalent description of this is the case where the transverse field is subject to Markovian, spatially uncorrelated classical noise. This is given by the master equation
\begin{align}
    \partial_t \hat \rho &= \gamma \sum_i \mathcal{D}[\hat \sigma_i^z] \hat \rho . \label{eqn:Znoise}
\end{align}
Upon performing the JW transformation, this maps to noise generated by the operators $\hat L_i  = \hat c_i^\dagger \hat c_i$. This is a known class of simulable fermionic models, as the jump operators are fermion bilinears and Hermitian, and does not require our technique to solve \cite{Shtanko2021}. 

Because the noise is quite simple in terms of the energy eigenmodes, the effect is has is quite simple. Specifically, it generates a length scale for \textit{neither the spins nor the fermions} as it is local in both representations. This simplicity allows one to observe that (see \cref{app:B.2}):
\begin{align}
    n_k(t) &= n_k(\infty)(1 - e^{-\gamma t}), 
\end{align}
where $n_k(\infty) \neq \frac{1}{2}$ now because the total $Z$ field is a conserved quantity and so the system does not equilibrate to infinite temperature. As one would expect, the fact that there is no length scale generated and the noise is local in the QP picture means that the phenomenology is quite different from the $XY$ noise previously examined. In \cref{fig:teff_v_noise}, one can see that there is never an emergent notion of temperature, as now the noise is truly agnostic to the spectrum of the Hamiltonian even at the critical point. 

We posit that, in some sense, the $Z$ noise is more fine-tuned than the $XY$ noise. For one thing, the $Z$-parity generates the $\Z_2$ symmetry of the Hamiltonian, and so the transverse field direction plays a special role. Moreover, one would imagine that in a general critical state with long range order, it would be difficult to construct local operators that couple only to QP bilinears with no higher powers.

{\subsection{Free fermion dissipation}}

The final type of noise we will consider is ``free fermion'' noise, where one simply drops the JW strings. This is evolution under the master equation
\begin{align}
    \partial_t \hat \rho &= \gamma \sum_i \mathcal{D}[\hat c_i] \hat \rho + \mathcal{D}[\hat c_i^\dagger] \hat \rho . \label{eqn:ffNoise}
\end{align}
This noise model is even simpler, as the entire Lindbladian is quadratic in the fermionic operators and Gaussian states remain Gaussian for all time. It is tempting to hope that this might be qualitatively similar to the original noise model \cref{eqn:XYnoise}. However, because the strings are responsible for the emergent length scale, the physical dynamics will be radically different. In fact, one can compute exactly that the quasiparticle density, for any value of $g$ in the initial condition, will evolve as
\begin{align}
    n_k(t) &= \frac{1}{2}(1 - e^{-2 \gamma t}) ,
\end{align}
which is completely independent of wavelength and energy. Similarly to the case of $Z$-dephasing noise, this will never develop a length scale or a temperature that is independent of energy. Even more surprisingly, under the free fermion dissipation the spins will actually generate a length scale. This can be seen immediately by noting that there is a duality relating spin observables under fermion noise to fermion observables under spin noise. 

\begin{figure}
    \centering
    \includegraphics{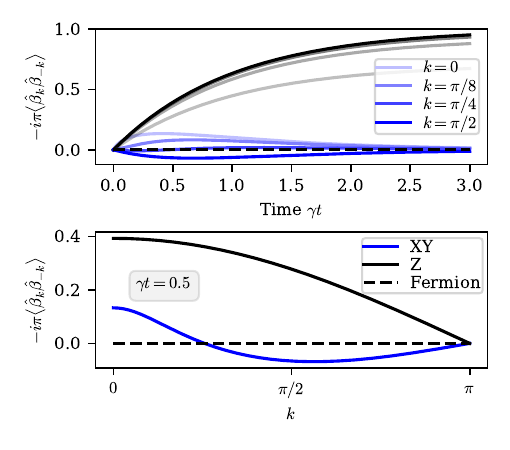}
    \caption{Quasiparticle coherences $C_{k,-k} = \langle \hat \beta_k \hat \beta_{-k} \rangle$ as a function of time (upper plot) and momentum (lower plot). The initial condition is the critical ground state $g/J = 1$. Blue curves show time evolution under the master equation \cref{eqn:XY_ME} with $\theta = 0$. Solid black curves show time evolution under the master equation \cref{eqn:Znoise} and dashed black curves show time evolution under the master equation \cref{eqn:ffNoise} }
    \label{fig:coherences}
\end{figure}

\bigskip
\subsection{Coherences}

In addition to the quasiparticle density, the only other two point correlation function allowed by translational invariance is the coherence $C_k = \langle \hat \beta_k \hat \beta_{-k} \rangle$. This parameter has three distinct behaviors for all three types of noise we have considered. Beginning with XY noise [c.f.~\cref{eqn:XYnoise}] we note that the steady state is infinite temperature, and so they must be identically zero at long times. Further, we find that they vary significantly with the value of $k$, and for $k = 0$ then $C^{XY}_{k=0}(t) = -\frac{i}{2\pi} (\gamma t) \log (\gamma t) + \mathcal{O}(\gamma t)$ grows superlinearly at short times. The exact functional form valid for all $(k,t)$ is derived explicitly in \cref{app:B.3}. We can compare this to the case where one drops the string operators, and instead considers free-fermion noise [c.f.~\cref{eqn:ffNoise}]. In this case, the coherences are identically zero for all $k$ and all times: $C_{k}^{\mathrm{ff}}(t) = 0$, in stark contrast to the spin noise. In this case, all of the generated coherences is a result of the interactions coming from the strings. Finally, as an intermediate case we can consider $Z$ dephasing [c.f.~\cref{eqn:Znoise}. Here, the steady state is no longer infinite temperature, and so the coherences actually saturate to a constant value: $C_k^Z(t) = \frac{i}{\pi}\mathrm{sgn}(k) \cos(k/2) (1- e^{-\gamma t})$, although the rate is $k$-independent. 

\bigskip
{\section{Critical XX Model}}
\label{sec:critXX}

\begin{figure}[t]
    \centering
    \includegraphics[width = \linewidth]{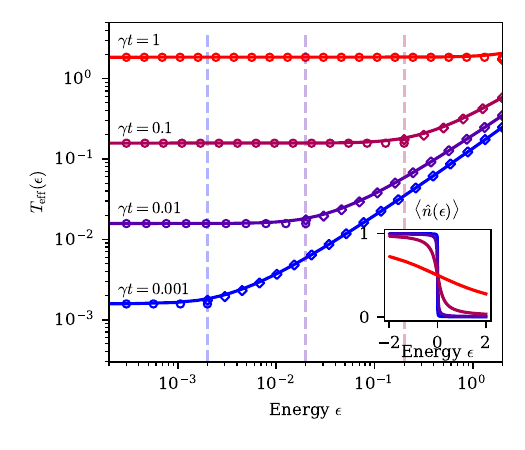}
    \caption{Effective temperature for the dephased XX model in the zero magnetization sector. Initial condition is the ground state, evolved under the XY noise channel for times $\gamma t \in \{ 0.001,0.01,0.1,1\}$. Solid lines correspond to the exact solution in \cref{eqn:XX_nk_exact}. Circle and diamond markers correspond to the long and short time expansions, respectively, given in \cref{eqn:TeffLimits_XX}. Vertical lines correspond to the cutoff for the approximation due to the length scale. Inset gives the quasiparticle densities.}
    \label{fig:XX_Temp}
\end{figure}

We show here that the main features found for a decohered critical ground state of the TFIM (namely the emergence of an effective temperature at low energies) can also arise in other models.  Specifically, we will now consider applying the same noise channel to an XX spin chain:
\begin{align}
    \hat H_{\mathrm{XX}} &= -\frac{J}{2} \sum_{i = 1}^{N-1} \hat \sigma_i^x \hat \sigma_{i + 1}^x + \hat \sigma_i^y \hat \sigma_{i + 1}^y,  \\
    \partial_t \hat \rho &= \gamma \mathcal{L}_{\mathrm{XY}} \hat \rho,  \label{eqn:XX_ME}
\end{align}
where $\mathcal{L}_{\mathrm{XY}}$ is defined in \cref{eqn:XYnoise}.  We will work with the ground state in the zero-magnetization sector. Note that while the CFT for the TFIM is given by free Majoranas with a central charge $c = 1/2$, for this XX model, the critical theory is the Luttinger liquid CFT with $c = 1$.  We show below that despite this difference in central charge, the physics of the emergent temperature in the decohered state is largely the same.

This Hamiltonian can be diagonalized in terms of the plane wave fermionic modes $\hat H_{\mathrm{XX}} = -2J\sum_k \cos(k) \hat d_k^\dagger \hat d_k$, see \cref{app:C} for more details.  As noted above, we work with th ground state in the zero magnetization sector, a state determined by: 
\begin{align}
    \langle \hat d_k^\dagger \hat d_k \rangle_\mathrm{GS} &= \left\{ \begin{array}{cc}
        1 & |k| < \pi/2  \\
        0 & |k| > \pi/2
    \end{array} \right. .
\end{align}
In real space, this state has critical long-range correlations given by [see \cref{app:C}]:
\begin{align}
    \langle \hat c_m^\dagger \hat c_n \rangle_\mathrm{GS} &= \frac{\sin \left( \frac{\pi}{2}(m-n) \right)}{\pi(m-n)}.
\end{align}
Using the new quasiparticle operators $n_k = \langle \hat n_k \rangle = \langle \hat d_k^\dagger \hat d_k \rangle$, we can again calculate explicitly
\begin{align}
    n_k(t) &= \frac{1}{2} + \frac{\arccot(e^{-i k + \gamma  t}) + \arccot (e^{i k + \gamma  t})}{\pi}. \label{eqn:XX_nk_exact}
\end{align}
From here, we can again calculate the short and long time expansions to find that
\begin{align}
    n_k(t) &= \left\{ 
    \begin{array}{cc}
       n_k(0) + \frac{2 J}{\pi \epsilon_k}\gamma t & \gamma t \lesssim |k|  \\
        \frac{1}{2} - \frac{\epsilon_k}{J \pi} e^{-\gamma t} & \gamma t \gtrsim |k| 
    \end{array}
    \right. .
\end{align}
Just as in the case with the TFIM, there is a kind of two stage relaxation: first the particle number grows very quickly at a rate inversely proportional to its energy. After this, every mode undergoes constant relaxation, but the initial jump has been imprinted as a kind of initial condition.

We can use these to calculate the temperature, and we again find that for long times/short wavevectors there is an emergent effective temperature constant in energy [see \cref{app:C}]
\begin{align}
    T_{\mathrm{eff}}(k,t) &\sim \left\{ \begin{array}{cc}
       \frac{\pi J}{2} \sinh(\gamma t)  & \gamma t \gtrsim |k|  \\
        |\epsilon_k| \log^{-1} \left( \frac{\pi |\epsilon_k|}{2J\gamma t}\right) & \gamma t \lesssim |k|    \end{array} \right.  . \label{eqn:TeffLimits_XX}
\end{align}
This scaling is confirmed in \cref{fig:XX_Temp}, where one can see qualitatively identical behavior to that seen in the TFIM model (compare to \cref{fig:effectiveTemp}).

Finally, we can also calculate how the overall quasiparticle changes at short times. Now, because we are considering the model in the zero-magnetization sector, the total quasiparticles number does not change in time. However, we can instead consider the imbalance between positive and negative energy particles. We will define 
\begin{align}
    n_{\epsilon>0} &= \int_{\pi/2}^\pi  \langle \hat d_k^\dagger \hat d_k \rangle \frac{\dd k}{\pi} + \int_{-\pi}^{-\pi/2}  \langle \hat d_k^\dagger \hat d_k \rangle \frac{\dd k}{\pi} ,
\end{align}
the particle density in positive energy modes. To leading order in time, this behaves as 
\begin{align}
     n_{\epsilon>0} &= -\frac{2}{\pi^2} \gamma t \log(\gamma t),
\end{align}
and so the particle number in positive energy modes is indeed growing superlinearly at short times, matching the TFIM, c.f. \cref{eqn:ntot_TFIM}. Interestingly, the prefactor differs by a factor of 2 which is the same factor by which the central charges differ.

All together, this strong provides evidence that the features identified for the TFIM and XX models could extend more generally to other critical states. 

\bigskip
{\section{Discussion}}
In this paper, we have developed a new tool for exactly and analytically solving the full time dynamics of any system that evolves under arbitrary matchgate circuits interleaved with arbitrary Pauli noise. We have used this technique to then analytically probe the time dynamics of a quantum critical state under spatially uniform, Markovian dephasing noise. We have shown that, despite the fact that the noise was infinite bandwidth and infinite temperature, there is an emergent notion of temperature at intermediate time scales coming from the interactions in the noisy dynamics.

This emergent temperature effect is intimately related to the fact that, even though the spins remain scale invariant, the underlying quasiparticle excitations described by the Jordan-Wigner fermions develop a time-dependent length scale. We confirmed that this behavior is not specific to the TFIM and extends to other decohered critical states in 1D. However, it would be extremely interesting to see if it persists in higher dimensional or non-integrable models.

Moreover, we have just begun to scratch the surface of the interplay between coherent dynamics and noise and there is still much to do in understanding how noise impedes unitary evolution in many-body systems near critical points. It would be extremely interesting to use the techniques developed here to try and gain either numerical or analytical insights into the effect of noise on the Kibble-Zurek mechanism or other adiabatic preparation mechanisms where one can see competition between unitary dynamics attempting to build long length scales while dissipation tries to shrink them. Such results could provide immediate insights into numerous noisy simulation experiments in the near term.

\bigskip
\begin{acknowledgments}
We would like to thank Amir Burshtein for reading an early draft of this manuscript and suggesting the critical XX model.
This work was supported by the Simons Foundation through a Simons Investigator award (Grant No. 669487).
\end{acknowledgments}

\bibliography{ref} 

\appendix
\onecolumngrid
\crefalias{section}{appendix}
\crefalias{subsection}{appendix}

\section{Exact Dynamics for matchgate circuits with Pauli noise}
\label{app:A}

We will be interested in both the continuous and discrete time evolution of a quantum state under matchgate circuits and local Pauli noise.
Matchgate circuits will be defined as any unitary transformation on a set of spins that can be written as the exponential of a quadratic, fermionic Hamiltonian. For example, take a one-dimensional spin model with the following Hamiltonian:
\begin{align}
    \hat H(\vec J, \vec{\tilde J}, \vec \Delta) &= \sum_j J_j \hat \sigma_j^+ \hat \sigma_{j + 1}^- + \tilde J_j \hat \sigma_j^+ \hat \sigma_{j + 1}^+ + \frac{\Delta_j}{2} \hat \sigma_j^z + \hc 
\end{align}
where the coefficients $J_j, \tilde J_j \in \mathbb{C}$ and $\Delta_j \in \R$ are arbitrary. Defining the set of canonical fermions
\begin{align}
    \hat c_j &= \left( \prod_{i < j} \hat \sigma_i^z \right) \hat \sigma_j^-,
\end{align}
then in terms of the fermionic degrees of freedom, the Hamiltonian can be rewritten as
\begin{align}
    \hat H(\vec J, \vec{\tilde J}, \vec \Delta) &= \sum_j J_j \hat c_j^\dagger \hat c_{j + 1} + \tilde J_j \hat c_j^\dagger \hat c_{j + 1}^\dagger + \Delta_j \hat c_j^\dagger \hat c_j + \hc 
\end{align}
and so any unitary 
\begin{align}
    \hat U_\theta(\vec J, \vec{\tilde J}, \vec \Delta) &= e^{i \theta \hat H(\vec J, \vec{\tilde J}, \vec \Delta)},
\end{align}
is an example of a matchgate. A matchgate circuit is one in which all unitary gates are compatible matchgates.  This is in no way an exhaustive list of all possible matchgate circuits, but it is one that is physically motivated (coming from the Hamiltonian of a 1D spin chain). Matchgate circuits can be efficiently simulated because they preserve the Gaussianity of fermionic states. That is to say, if the state begins as a Gaussian fermionic state, then it will remain a Gaussian fermionic state for all times under the circuit evolution. A Gaussian state is completely described by its covariance matrix. If one defines a set of Majorana modes
\begin{align}
    \hat \gamma_{2m} = \hat c_m + \hat c_m^\dagger, \ \ \ \ \ \ \hat \gamma_{2m + 1} = i(\hat c_m - \hat c_m^\dagger), \label{eqn:Maj}
\end{align}
then the covariance matrix is $\Gamma_{m,n} = \langle \hat \gamma_m \hat \gamma_n\rangle$. This is an antisymmetric matrix described by $N(2N-1)$ real coefficients if there are $N$ spins in the problem. The dynamics of the covariance matrix can be computed in polynomial time, and so the dynamics are efficiently classically simulable. Interspersed in the matchgate circuit, we also consider Pauli noise. Pauli noise acts on the local qubit degrees of freedom via the Kraus channels:
\begin{align}
    \mathcal{K}[\hat \rho] &= \sum_{s \in \{0,1\}^{2n}} p_s \hat K_s \hat \rho \hat K_s^\dagger,  \\
    \hat K_s &= \sqrt{p_s} \prod_{i = 0}^{n-1} (\hat \sigma_i^x)^{s_{2i}} (\hat \sigma_i^z)^{s_{2i + 1}},
\end{align}
where $\sum_s p_s = 1$.

At this point, we note that the given dynamics falls within the class of stochastically simulable dynamics given in Ref.~\cite{Pocklington2025}. This result implies that, while the Pauli noise does not preserve the Gaussianity of the density matrix, it maps the state to a convex sum of Gaussian states which can be efficiently sampled from to compute observables to fixed additive error. However, the special structure of the Pauli noise allows us to go one step further, and exactly re-average the stochastic equation of motion. To understand this, let's consider a single time step of first applying the matchgate circuit followed by the noise. We have that
\begin{align}
    \hat \rho \mapsto \mathcal{M}[\hat \rho] \equiv \sum_s  \hat K_s \hat U \hat \rho \hat U^\dagger \hat K_s^\dagger = \sum_s \tr(\hat K_s \hat U \hat \rho \hat U^\dagger \hat K_s^\dagger) \frac{\hat K_s \hat U \hat \rho \hat U^\dagger \hat K_s^\dagger}{\tr(\hat K_s \hat U \hat \rho \hat U^\dagger \hat K_s^\dagger)} = \sum_s  N_s \hat \rho_s.
\end{align}
Here, we have defined the normalized states $\hat \rho_s$ and the normalization constants $N_s$ which can depend on the state. For any fixed bitstring $s$, then $\hat K_s U$ maps Gaussian states to Gaussian states, and so it has an induced action on the set of fermionic covariance matrices. Let us define this implicit transformation by considering a Gaussian state $\hat \rho_\Gamma$ with covariance matix $\Gamma$:
\begin{align}
    \Gamma_{mn} &= \tr( \hat \rho_\Gamma \hat \gamma_m \hat \gamma_n ) \mapsto \sum_s  N_s(\Gamma) \tr(\hat \rho_s(\Gamma) \hat \gamma_m \hat \gamma_n ) \equiv \sum_s  N_s(\Gamma) [\mathcal{K}_{s,\mathrm{cov}}(\Gamma)]_{mn},
\end{align}
where the subscript ``cov'' denotes a transformation on the space of covariance matrices. Therefore, after a single time step, we have that
\begin{align}
    \Gamma \mapsto \mathcal{M}_\mathrm{cov} [\Gamma] &= \sum_s p_s N_s \mathcal{K}_{s,\mathrm{cov}}[\Gamma] = \overline{\mathcal{K}}_\mathrm{cov}[\Gamma],
\end{align}
where the overline denotes averaging with respect to the probability distribution $\{ N_s \}$. To understand the dynamics up to a time step $t$ requires one to compute
\begin{align}
    \mathcal{M}_\mathrm{cov}^t &= \overline{\mathcal{K}^t_\mathrm{cov}}.
\end{align}
Note that $\mathcal{M}_\mathrm{cov}$ is simple to compute, but the difficulty in computing its powers need not necessarily be simple if there are strong correlations different time steps. In general, computing powers of $\mathcal{M}$ can be done by stochastically sampling over trajectories (as is done in Ref.~\cite{Pocklington2025}). However, here is where Pauli noise becomes useful: because the channel is a convex sum of unitary transformations, we have that $N_s(\Gamma) = \xi_s$ is \textit{independent of the state}. Therefore, there are no correlations between different time steps: i.e., $\overline{\mathcal{K}_\mathrm{cov}^t} = (\overline{\mathcal{K}_\mathrm{cov}})^t$, and so directly re-averaging the stochastic equations of motion become trivial.

Note that this is true for \textit{arbitrary} Pauli strings, as for any Pauli string $\hat P$ we have that $\hat P^\dagger \hat P = 1$, and further that every Pauli string under JW maps to a Majorana string, which is a Gaussian unitary operator. Going forward, we will focus primarily on \textit{local} Pauli channels as they are often the most physically relevant, but the results all hold for arbitrary Pauli channels.

\subsection{Discrete time dynamics}
\label{app:A.1}

Here, we will directly compute the evolution of the two point functions under local Pauli channels. For simplicity, we will assume the noise rates are spatially uniform, but this is not required and can be relaxed in an extremely straightforward manner. 

The Kraus channel for each different type of Pauli noise are commuting, so we can calculate each of them independently. We will do this in two different methods. The first way is the one inspired by Ref.~\cite{Pocklington2025}, where we consider the action of the strings on the fermionic two-point functions. Let's begin by defining the sign variables:
\begin{align}
    \eta_{i,j} &= \left\{ 
    \begin{array}{cc}
       1  &  i < j \\
       -1  & i \geq j
    \end{array}
    \right. .
\end{align}
Recall the canonical Majorana modes defined in \cref{eqn:Maj}. Then we have that 
\begin{align}
    \hat \sigma_j^x &= \left( \prod_{i < j} (-1)^{\hat n_i } \right) (\hat c_j + \hat c_j^\dagger) = \left( \prod_{i < j} -i\hat \gamma_{2i} \hat \gamma_{2i + 1} \right) \hat \gamma_{2j}, \\
    \hat \sigma_j^y &= \left( \prod_{i < j} (-1)^{\hat n_i } \right) i (\hat c_j - \hat c_j^\dagger) = \left( \prod_{i < j} -i\hat \gamma_{2i} \hat \gamma_{2i + 1} \right) \hat \gamma_{2j + 1}, \\
    \hat \sigma_j^z &= 2 \hat c_j^\dagger \hat c_j - 1 = i \hat \gamma_{2j} \hat \gamma_{2j + 1}.
\end{align}
Clearly, $z$-noise is the simplest, as it does not have JW strings. If we define the $z$-type channel $\mathcal{K}_{i,z}$ to be the induced covariance matrix transformation under $\hat \rho \mapsto p_z \hat \sigma_i^z \hat \rho \hat \sigma_i^z + (1 - p_z) \hat \rho$ and then define $\mathcal{K}_z = \prod_i \mathcal{K}_{i,z}$ we find that
\begin{align}
    (\mathcal{K}_{j,z,\mathrm{cov}}[\Gamma])_{mn} &= p_z \langle (i \hat \gamma_{2j} \hat \gamma_{2j + 1}) \hat \gamma_m \hat \gamma_n (i \hat \gamma_{2j} \hat \gamma_{2j + 1} ) \rangle_{\Gamma} + (1 - p_z) \Gamma_{mn} \nonumber  \\
    &= \left( p_z \left[ 4 \delta_{2j,m} \delta_{2j + 1,n} + 4  \delta_{2j,n} \delta_{2j + 1,m} - 2 \delta_{2j,m} - 2 \delta_{2j + 1,m} - 2 \delta_{2j,n} - 2 \delta_{2j + 1,n} + 1 \right] + (1 - p_z) \right) \Gamma_{mn}. \\
    \implies (\mathcal{K}_{z,\mathrm{cov}}[\Gamma])_{mn} &= \Gamma_{mn} \times \left\{ \begin{array}{cc}
      1   &  \lfloor m/2 \rfloor = \lfloor n/2 \rfloor\\
      (1 - 2p_z)^2   &  \mathrm{otherwise}
    \end{array} 
    \right. .
\end{align}
For $x$- and $y$- type noise, we will see extremely different behavior. Consider $x$-noise now, given by 
\begin{align}
     (\mathcal{K}_{j,x,\mathrm{cov}}[\Gamma])_{mn} &= p_x \left\langle \hat \gamma_{2j}  \left( \prod_{i < j} -i\hat \gamma_{2i} \hat \gamma_{2i + 1} \right)  \hat \gamma_m \hat \gamma_n \left( \prod_{i < j} -i\hat \gamma_{2i} \hat \gamma_{2i + 1} \right) \hat \gamma_{2j}  \right\rangle_{\Gamma} + (1 - p_x) \Gamma_{mn} \nonumber \\
     &= p_x \eta_{j,\floor{m/2}} \eta_{j,\floor{n/2}} \langle \hat \gamma_{2j} \hat \gamma_m \hat \gamma_n \hat \gamma_{2j} \rangle_\Gamma + (1 - p_x) \Gamma_{mn} \nonumber \\
     &= \left[ p_x \eta_{j,\floor{m/2}} \eta_{j,\floor{n/2}} (1 - 2 \delta_{2j,m} - \delta_{2j,n}) + (1 - p_x) \right] \Gamma_{mn}. 
\end{align}
If we assume that $m < n$ (the remaining covariances can be found using the skew-symmetry of the covariance matrix), and define $m_2 = m \mod 2$ and vice versa for $n_2$, then:
\begin{align}
    (\mathcal{K}_{x,\mathrm{cov}}[\Gamma])_{mn} &= \Gamma_{mn} \times 
    \left\{
    \begin{array}{cc}
      (1 - 2 p_x)^{|\floor{m/2}-\floor{n/2}| - m_2 + n_2}   & |\floor{m/2}-\floor{n/2}| > 1 \\
       (1 - 2 p_x)  & |\floor{m/2}-\floor{n/2}| = 0
    \end{array}
    \right. .
\end{align}
A similar calculation gives that
\begin{align}
    (\mathcal{K}_{y,\mathrm{cov}}[\Gamma])_{mn} &= \Gamma_{mn} \times     \left\{
    \begin{array}{cc}
      (1 - 2 p_y)^{|\floor{m/2}-\floor{n/2}| + m_2 - n_2}   & |\floor{m/2}-\floor{n/2}| > 1 \\
       (1 - 2 p_y)  & |\floor{m/2}-\floor{n/2}| = 0
    \end{array}
    \right. .
\end{align}
Putting these results together, for example if we take an infinite temperature reservoir with Kraus channels given by equal probability of exciting or de-exciting each spin (which is modeled by equal decoherence in the $X$ and $Y$ direction) as considered in the main text, we find that
\begin{align}
    (\mathcal{K}_{x,\mathrm{cov}}[\mathcal{K}_{y,\mathrm{cov}}[\Gamma]])_{mn} &= \Gamma_{mn} \times     \left\{
    \begin{array}{cc}
      (1 - 2 p)^{2|\floor{m/2}-\floor{n/2}|}   & |\floor{m/2}-\floor{n/2}| > 1 \\
       (1 - 2 p)^{2(1 - \delta_{mn})}  & |\floor{m/2}-\floor{n/2}| = 0
    \end{array}
    \right. ,
\end{align}
where $p = p_x = p_y$. For complete depolarizing noise, we have that
\begin{align}
    (\mathcal{K}_{z,\mathrm{cov}}[\mathcal{K}_{x,\mathrm{cov}}[\mathcal{K}_{y,\mathrm{cov}}[\Gamma]]])_{mn} &= \Gamma_{mn} \times  (1 - 2 p)^{2(|\floor{m/2}-\floor{n/2}| + 1 - \delta_{mn}) } ,
\end{align}
where $p = p_x = p_y = p_z$.

A much simpler way to compute the two point functions is to, instead of working in terms of the fermion variables, note that they can be written in terms of Pauli strings. We have that (for $m < n$)
\begin{subequations}
\begin{align}
    \Gamma_{2m, 2n} &= i\langle \hat Y_m \hat Z_{m + 1} \dots \hat Z_{n -1} \hat X_n \rangle, \\
    \Gamma_{2m, 2n + 1} &= i\langle \hat Y_m \hat Z_{m + 1} \dots \hat Z_{n -1} \hat Y_n \rangle, \\
    \Gamma_{2m + 1, 2n} &= i\langle \hat X_m \hat Z_{m + 1} \dots \hat Z_{n -1} \hat X_n \rangle, \\
    \Gamma_{2m + 1, 2n + 1} &= i\langle \hat X_m \hat Z_{m + 1} \dots \hat Z_{n -1} \hat Y_n \rangle .
\end{align}
\end{subequations}
From here, it is simple to simply note that Pauli strings are mapped to themselves under Pauli channels, and then count the number operators that anticommute to get the correct prefactors. This completely recovers the previous result, though does not immediately elucidate the convex sum of Gaussians structure.

\subsection{Continuous time dynamics}
\label{app:A.2}

We can similarly define the continuous time dynamics one would obtain by considering, for example, the limit $p_{x,y,z} = \gamma_{x,y,z} \dd t$ as $\dd t \to 0$. This gives Lindblad dynamics
\begin{align}
    \partial_t \hat \rho &= \sum_i \sum_{\alpha \in \{x,y,z\} } \gamma_\alpha \mathcal{D}[\hat \sigma_i^\alpha] \hat \rho .
\end{align}
This induces the dynamics on the covariance matrix
\begin{align}
    \partial_t \Gamma_{mn} &= -(\gamma_1 |\floor{m/2} - \floor{n/2}| + \gamma_2) \Gamma_{mn} , \\
    \gamma_1 &= \gamma_x + \gamma_y , \\
    \gamma_2 &= (1 - \delta_{\floor{m/2}, \floor{n/2}})(2\gamma_z + (\gamma_x - \gamma_y)(n_2 - m_2))  + \delta_{\floor{m/2}, \floor{n/2}} (1 - \delta_{mn})(\gamma_x + \gamma_y) ,
\end{align}
where we have re-used the variables $m_2 = m\mod 2, n_2 = n \mod 2$. Note again that if $\gamma_x = \gamma_y = \gamma_z = \gamma$, we have the simplification $\gamma_1 = 2 \gamma, \gamma_2 = 2 \gamma (1 - \delta_{mn})$.

\subsection{Non-Gaussianity and violation of Wick's theorem}
\label{app:A.3}

Here, we show that the state is able to massively violate Wick's theorem despite the fact that the dynamics of moments close on themselves. To see this, let's try and calculate the connected correlation function $\langle \langle \hat \sigma_m^z \hat \sigma_n^z \rangle \rangle \equiv \langle \hat \sigma_m^z \hat \sigma_n^z \rangle - \langle \hat \sigma_m^z \rangle \langle  \hat \sigma_n^z \rangle$ in two ways, once using the full dynamics, and once using Wick's theorem, and compare the two. Under the XY noise channel, this observable (being a Pauli) evolves extremely simply and the dynamics are described by
\begin{align}
    \langle \langle \hat \sigma_m^z \hat \sigma_n^z \rangle \rangle(t) &= e^{-2 \gamma t} \langle \langle \hat \sigma_m^z \hat \sigma_n^z \rangle \rangle(0).
\end{align}
Alternatively, if we instead were to pretend that the state was Gaussian for all times, then we could use Wick's theorem, and ask how well we do in predicting observables. This would give us
\begin{align}
     \langle \langle \hat \sigma_m^z \hat \sigma_n^z \rangle \rangle_W(t) &= -\langle  \hat \gamma_{2m} \hat \gamma_{2m + 1} \hat \gamma_{2n} \hat \gamma_{2n + 1} \rangle_W(t) + \langle  \hat \gamma_{2m} \hat \gamma_{2m + 1} \rangle (t) \langle  \hat \gamma_{2n} \hat \gamma_{2n + 1} \rangle(t) \nonumber  \\
     &= \langle \hat \gamma_{2m} \hat \gamma_{2n} \rangle(t) \langle  \hat \gamma_{2m + 1} \hat \gamma_{2n + 1} \rangle(t)  - \langle \hat \gamma_{2m} \hat \gamma_{2n + 1} \rangle(t) \langle  \hat \gamma_{2m + 1} \hat \gamma_{2n} \rangle(t) \nonumber \\
     &= e^{-4 \gamma |m-n| t} \langle \langle \hat \sigma_m^z \hat \sigma_n^z \rangle \rangle(0),
\end{align}
where the subscript $W$ denotes we use Wick's theorem to evaluate the 4 point correlation functions. Note that this differs \textit{exponentially} from the actual result, and so Wick's theorem fails catastrophically to evaluate the dynamics for lattice sites with large spatial separation.

To understand intuitively what is going on, it is useful to consider again the exact form of the density matrix $\hat \rho$ as a function of time. We can write $\hat \rho$ as
\begin{align}
    \hat \rho &= \sum_{\vec s \in \{0,1\}^{2N} } p_{s} |\Gamma_{\vec s} \rangle \langle \Gamma_{\vec s} |, \\
    p_s &= p^s(1 - p)^s,  \ \ \ \ p = \frac{1}{2}(1 - e^{-\gamma t/2}).
\end{align}
Here, $p_{\vec s}$ is a classical probability distribution, $s = |\vec s|_1$ its 1-norm, and $|\Gamma_{\vec s} \rangle$ is a Gaussian state with a covariance matrix $\Gamma_{\vec s}$. This covariance matrix is defined by
\begin{align}
    \Gamma_{\vec s} &= \left( \prod_{i = 1}^{2N} U_i^{s_i} \right) \Gamma_0 \left( \prod_{i = 1}^{2N} U_i^{s_i} \right)^\dagger , \\
    (U_{2i})_{mn} &= \delta_{mn} (2 \Theta(m-2i + 0.5) - 1) , \\
    (U_{2i + 1})_{mn} &= \delta_{mn} (2 \Theta(m-2i - 0.5) - 2 \delta_{m,2i+ 2} - 1),
\end{align}
where $\Gamma_0$ is the initial fermionic covariance matrix in terms of the Majorana modes. Essentially, each matrix $U_{2i}/U_{2i + 1}$ multiplies the coherences between operators on either side of the lattice site $i$ by a sign. Now, when two lattice sites are very far away, there are many locations one can insert minus signs, giving the emergent length scale. However, note that when we look at something that depends quadratically on the covariance matrix (e.g. the 4 point functions), then all of the minus signs cancel out. We can see this in action by once more calculating the connected correlation function, but this time use the actual representation of the fermionic state. We wish to know
\begin{align}
    \tr(\hat \gamma_{2m} \hat \gamma_{2m + 1} \hat \gamma_{2n} \hat \gamma_{2n + 1} \hat \rho) &= \sum_{\vec s \in \{0,1\}^{2N} } p_{s}  \langle \Gamma_{\vec s} | \hat \gamma_{2m} \hat \gamma_{2m + 1} \hat \gamma_{2n} \hat \gamma_{2n + 1} |\Gamma_{\vec s} \rangle \nonumber \\
    &= \sum_{\vec s \in \{0,1\}^{2N} } p_{s} \left[ (\Gamma_{\vec s})_{2m,2m + 1}(\Gamma_{\vec s})_{2n,2n + 1} + (\Gamma_{\vec s})_{2m,2n + 1}(\Gamma_{\vec s})_{2m + 1,2n} - (\Gamma_{\vec s})_{2m,2n}(\Gamma_{\vec s})_{2m + 1,2n + 1}\right] .
\end{align}
From here, we note that the \textit{product} of two covariance matrix elements only picks up a sign if exactly one of them did. The only way this occurs is if, considering only the four elements $s_{2m}, s_{2m + 1}, s_{2n}, s_{2n + 1}$, then an odd number of them are positive. Since these are the only relevant terms, we can simplify the problem significantly to:
\begin{align}
     \tr(\hat \gamma_{2m} \hat \gamma_{2m + 1} \hat \gamma_{2n} \hat \gamma_{2n + 1} \hat \rho) &= \sum_{\vec s \in \{0,1\}^{4} } p_{s} \left[ (\Gamma_{\vec s})_{2m,2m + 1}(\Gamma_{\vec s})_{2n,2n + 1} + (\Gamma_{\vec s})_{2m,2n + 1}(\Gamma_{\vec s})_{2m + 1,2n} - (\Gamma_{\vec s})_{2m,2n}(\Gamma_{\vec s})_{2m + 1,2n + 1}\right] \nonumber \\
     &= \left( \sum_{\vec s \in \{0,1\}^{4}, |\vec s|_2 = 0 } p_{s} - \sum_{\vec s \in \{0,1\}^{4}, |\vec s|_2  = 1 } p_{s}\right) \langle \hat \gamma_{2m} \hat \gamma_{2m + 1} \hat \gamma_{2n} \hat \gamma_{2n + 1} \rangle_{t = 0} \nonumber \\
     &= e^{-2 \gamma t} \langle \hat \gamma_{2m} \hat \gamma_{2m + 1} \hat \gamma_{2n} \hat \gamma_{2n + 1} \rangle_{t = 0} , 
\end{align}
where we have defined $|\vec s|_2 $ to be the Hamming weight of the bitstring $\vec s$ modulo 2. We can clearly see that the fact that the covariances matrices add incoherently means that, while there can destructive interference when calculating the average two-point functions, it can also go away when calculating higher point correlators and so Wick's theorem tends to not hold even approximately.

\subsection{Periodic boundary conditions}
\label{app:A.4}

\begin{figure}
    \centering
    \includegraphics{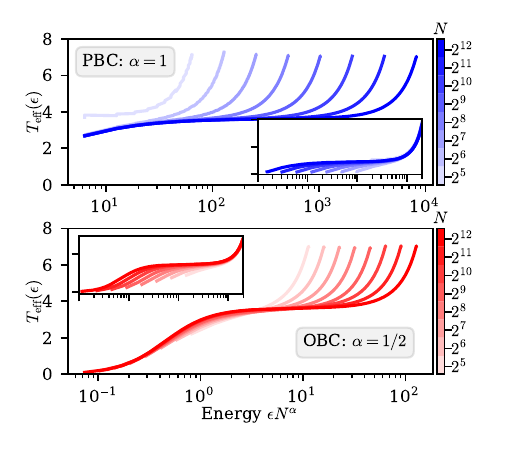}
    \caption{Effective temperature for both periodic (upper) and open (lower) boundary conditions at the critical point with $J = g = \gamma  = 1$ at time $t = 1$. We consider system sizes from $N = 32$ to $N = 4096$. The horizontal energy axis is rescaled by $N^\alpha$, where $\alpha = 1/2$ for open boundary conditions and $\alpha = 1$ for closed boundary conditions. Inset: same data as main plot without rescaling ($\alpha = 0$). }
    \label{fig:bc_sclaing_collapse}
\end{figure}

There is some subtlety that comes from periodic boundary conditions, as the Hamiltonian does not map to a completely free model under the JW transformation. The problem lies in the fact that to connect the lattice sites $1$ and $N$ picks up a JW string equivalent to the total parity of the state:
\begin{align}
    \hat H_{\mathrm{PBC}}(\vec J, \vec{\tilde J}, \vec \Delta) &= J_N \hat \sigma_N^+ \hat \sigma_{1}^- + \tilde J_N \hat \sigma_N^+ \hat \sigma_{1}^+ + \sum_{j = 1}^{N-1} J_j \hat \sigma_j^+ \hat \sigma_{j + 1}^- + \tilde J_j \hat \sigma_j^+ \hat \sigma_{j + 1}^+ + \sum_{j = 1}^{N}\frac{\Delta_j}{2} \hat \sigma_j^z + \hc 
\end{align}
Now, when doing JW, this becomes:
\begin{align}
    \hat H_{\mathrm{PBC}}(\vec J, \vec{\tilde J}, \vec \Delta) &= -\left[ J_N \hat c_N^\dagger \hat c_1^- + \tilde J_N \hat c_N^\dagger \hat c_1^\dagger \right] \hat \Pi + \sum_{j=1}^{N-1} J_j \hat c_j^\dagger \hat c_{j + 1} + \tilde J_j \hat c_j^\dagger \hat c_{j + 1}^\dagger + \sum_{j=1}^{N} \Delta_j \hat c_j^\dagger \hat c_j + \hc 
\end{align}
where we have defined the total parity operator $\hat \Pi = \prod_{j = 1}^N (-1)^{\hat n_j}$. Recall that $\hat \Pi$ is the generator of a weak $\Z_2$ symmetry (as fermions must obey superselection rules), and so we have that
\begin{align}
    [ \hat H_{\mathrm{PBC}}, \hat \Pi] &= 0 .
\end{align}
This means that the Hamiltonian can be decomposed as a direct sum of $\hat \Pi$ eigenspaces
\begin{align}
    \hat H_{\mathrm{PBC}}^{\pm} &= \mp \left[ J_N \hat c_N^\dagger \hat c_1^- + \tilde J_N \hat c_N^\dagger \hat c_1^\dagger \right] + \sum_{j=1}^{N-1} J_j \hat c_j^\dagger \hat c_{j + 1} + \tilde J_j \hat c_j^\dagger \hat c_{j + 1}^\dagger + \sum_{j=1}^{N} \Delta_j \hat c_j^\dagger \hat c_j + \hc ,
\end{align}
where the parity is replaced with its expectation value. Now, both $\hat H_{\mathrm{PBC}}^\pm$ are quadratic, and so their dynamics can be efficiently simulated. For a state with positive parity, one should evolve it under $\hat H_{\mathrm{PBC}}^+$ and vice-versa for odd parity states. Now, we note that the jumps always map a state with fixed parity to another state with fixed parity, as 
\begin{align}
    \hat \Pi \hat \sigma_i^\alpha &= \lambda_\alpha \hat \sigma_i^\alpha \hat \Pi  , \ \ \ \lambda_x = \lambda_y = -1, \lambda_z = 1 .
\end{align}
Therefore, it tells us that under these dynamics the state will always be an incoherent mixture of different parity sectors. Thus, one can write the state as
\begin{align}
    \hat \rho &= \hat \rho_+ + \hat \rho_- = \sum_{\Gamma_+} p(\Gamma_+) |\Gamma_+ \rangle \langle \Gamma_+ | + \sum_{\Gamma_-} p(\Gamma_-) |\Gamma_- \rangle \langle \Gamma_- | , 
\end{align}
where subscripts $+,-$ refer to even or odd states, respectively. The density matrices $\hat \rho_+ , \hat \rho_- $ are each independently convex sums of either even or odd pure Gaussian states described by covariances $\Gamma_{+,-}$. The total covariance matrix of the state $\hat \rho$ is given by
\begin{align}
    \tr( \hat \rho \hat \gamma_{m} \hat \gamma_n) &= \sum_{\Gamma_+} p(\Gamma_+) (\Gamma_+)_{m,n} + \sum_{\Gamma_-} p(\Gamma_-) (\Gamma_-)_{m,n}
\end{align}
and similarly for higher moments. As for the dynamics, only the quantum jumps can change the parity of the state, and so the coherent dynamics are extremely simple:
\begin{align}
    \partial_t \hat \rho_{\pm} &= -i[\hat H^\pm_{\mathrm{PBC}}, \hat \rho_{\pm}]
\end{align}
The noise, though, will mix the sectors in an incoherent fashion:
\begin{align}
    \partial_t \langle  \hat \gamma_m \hat \gamma_n \rangle_{\pm} &= \sum_{\alpha = x,y} \sum_{j = 1}^N \gamma_\alpha \left( \langle \hat \sigma_j^\alpha \hat \gamma_m \hat \gamma_n \hat \sigma_j^\alpha \rangle_{\mp} - \langle \hat  \gamma_m \hat \gamma_n  \rangle_{\pm} \right) + \sum_{j = 1}^N \gamma_z \left( \langle \hat \sigma_j^z\hat \gamma_m \hat \gamma_n \hat \sigma_j^z \rangle_{\pm} - \langle \hat  \gamma_m \hat \gamma_n  \rangle_{\pm} \right)
\end{align}
where subscripts refer to taking expectation values with respect to either the even or odd parity sectors. What this means is that by keeping track of two different covariance matrices, one can evolve the system with periodic boundary conditions. In principle, this result is more general than just PBC. Anytime there is a model with a weak symmetry generator $\hat G$, and the Hamiltonian is interacting in the form $\hat H = \hat H_0 + \hat H_1 \hat G $ where both $\hat H_0$ and $\hat H_1$ are free, then this would work, where one needs to keep track of as many covariance matrices as there are eigenspaces of $\hat G$.

In \cref{fig:bc_sclaing_collapse}, we recreate the data in \cref{fig:Hamiltonian Data} for both open and periodic boundary conditions at $\gamma t = 1$. We find that while periodic boundary conditions converge at low energies like $1/N$, for periodic boundary conditions it only converges at $1/\sqrt{N}$, and so there is value in using periodic boundary conditions as it converges significantly more quickly with large system sizes.

\section{Ising Model Results}
\label{app:B}

To set the stage, let's define the Ising model Hamiltonian as
\begin{align}
    \hat H &= \sum_i -J \hat \sigma_i^x \hat \sigma_{i + 1}^x +  g \hat \sigma_i^z.
\end{align}
The Hamiltonian can be rewritten as a quadratic form of the free fermion modes:
\begin{align}
     \hat H &= \sum_i -J (\hat \sigma_i^+ + \hat \sigma_i^-)(\hat \sigma_{i + 1}^+ + \hat \sigma_{i + 1}^-) +  g (2 \hat \sigma_i^+ \hat \sigma_i^- - 1) = \sum_i -J (\hat c_i^\dagger - \hat c_i ) (\hat c_{i + 1}^\dagger + \hat c_{i + 1}) +  g (2 \hat c_i^\dagger \hat c_i  - 1).
\end{align}
If we now define the set of Fourier modes $\hat d_k = \sum_j e^{ijk} \hat c_j$, then this becomes
\begin{align}
    \hat H &= \sum_{k} -2J \cos(k) \hat d_k^\dagger \hat d_k + i J \sin(k) (\hat d_k \hat d_{-k} + \hat d_k^\dagger \hat d_{-k}^\dagger) +  g (2 \hat d_k^\dagger \hat d_k  - 1).
\end{align}
This is almost diagonal, up to the fact that there is mixing between modes at $\pm k$. To solve this, we will implement a Bogoliubov transformation defined via the modes 
\begin{align}
    \hat \beta_k &= \cos(\theta_k/2) \hat d_k + i \sin(\theta_k/2) \hat d_{-k}^\dagger.
\end{align}
Note that this implies 
\begin{align}
    \sum_k \epsilon_k \hat \beta_k^\dagger \hat \beta_k &= \sum_k \epsilon_k \left[ \cos(\theta_k) \hat d_k^\dagger \hat d_k + \sin^2(\theta_k/2) + i/2 \sin(\theta_k) (\hat d_k^\dagger \hat d_{-k}^\dagger - \hat d_{-k} \hat d_k)  \right].
\end{align}
From here, we can directly read out that
\begin{align}
    \epsilon_k \cos(\theta_k) &=  -2J \cos(k) + 2 g, \ \ \ \ \ \epsilon_k \sin(\theta_k) = 2J \sin(k). \\
    \implies \epsilon_k &= 2 J \sqrt{( \cos(k) - g/J)^2 +  \sin^2(k)}, \\
    \implies \theta_k &= \arctan \left( \frac{ \sin(k)}{ g/J  -\cos(k)} \right),
\end{align}
as discussed in the main text. Here and henceforth, we will work in units where $J=1$, so the phase transition occurs at $g = 1$. We can use this form to calculate the ground state expectation value of observables by noting that, since every mode is defined as having positive energy they are all in vacuum in the ground state. Hence
\begin{align}
    \langle \hat c_m^\dagger \hat c_n \rangle_{\mathrm{GS}} &= \frac{1}{N} \sum_k e^{ik(m-n))} \langle \hat d_k^\dagger \hat d_k \rangle \nonumber \\
    &=  \frac{1}{N} \sum_k e^{ik(m-n))} \langle (\cos(\theta_k/2) \hat \beta_k^\dagger + i \sin(\theta_k/2) \hat \beta_{-k} ) (\cos(\theta_k/2) \hat \beta_k -i \sin(\theta_k/2) \hat \beta_{-k}^\dagger ) \rangle \nonumber  \\
    &= \frac{1}{N} \sum_k e^{ik(m-n)} \sin^2(\theta_k/2) = \frac{1}{2} \delta_{mn} -  \frac{1}{2N} \sum_k e^{ik(m-n)} \cos(\theta_k) = \frac{1}{2} \delta_{mn} -  \frac{1}{2} \mathcal{F}_{\cos \theta}(m-n),
\end{align}
where we have defined the Fourier transform of $\cos(\theta_k)$ to be $\mathcal{F}_{\cos \theta}$. A similar calculation tells us that
\begin{align}
    \langle \hat c_m \hat c_n \rangle_\mathrm{GS} &= -\frac{i}{2} \mathcal{F}_{\sin \theta}(m-n).
\end{align}
Exactly at the phase transition, these can be evaulated exactly as the form of $\theta_k $ becomes significantly simpler. When $g = J$,
\begin{align}
    \theta_k &= \left\{ \begin{array}{cc}
        \frac{\pi}{2} - \frac{k}{2} & 0 < k < \pi  \\
        - \frac{\pi}{2} - \frac{k}{2} & -\pi < k < 0  \\
    \end{array}
    \right. .
\end{align}
This gives us the following results
\begin{align}
    \mathcal{F}_{\cos \theta}(d) &= \frac{1}{N} \sum_k e^{ikd} \cos(\theta_k) = \frac{1}{N} \sum_{k} e^{ikd} |\sin(k/2)| \sim \frac{1}{2\pi} \int_{-\pi}^\pi e^{ikd} |\sin(k/2)| = \frac{1}{\pi} \frac{2}{1 - 4d^2}, \label{eqn:cor1} \\
    \mathcal{F}_{\sin \theta}(d) &= \frac{1}{N} \sum_k e^{ikd} \sin(\theta_k) = \frac{1}{N} \sum_k e^{ikd} \mathrm{sgn}(k) \cos(k/2)  \sim \frac{1}{2\pi} \int_{-\pi}^\pi e^{ikd} \mathrm{sgn}(k) \cos(k/2) = -\frac{i}{\pi} \frac{4 d}{1 - 4d^2}, \label{eqn:cor2}
\end{align}
which gives, as expected, algebraically decaying correlations at the phase transition. It is worthwhile to note that the cause of the algebraic decay is exactly due to the non-analytic behavior of $\theta_k$ at the phase transition. When it behaves smoothly (and similarly $\cos(\theta),\sin(\theta)$ are smooth), then the correlations functions are the Fourier transform of a smooth function which necessarily must decay faster than any polynomial. 

\subsection{Quasiparticle number in time}
\label{app:B.1}

Here, we will derive the full time dynamics for the energy-resolved quasiparticle numbers in time under the noise. Using the results from the XY channel derived earlier, we have that
\begin{align}
    \langle \hat c_m^\dagger \hat c_n \rangle(t) &= e^{-\gamma |m-n|t - \delta_{mn} \gamma t} \langle \hat c_m^\dagger \hat c_n \rangle(0) + \frac{1}{2} \delta_{mn} (1 - e^{-\gamma t}) \nonumber  \\
    &= e^{-\gamma |m-n|t - \delta_{mn} \gamma t} \left( \frac{1}{2} \delta_{mn} -  \frac{1}{2} \mathcal{F}_{\cos \theta}(m-n) \right) + \frac{1}{2} \delta_{mn} (1 - e^{-\gamma t}) \nonumber \\
    &= -\frac{1}{2} e^{-\gamma |m-n|t - \delta_{mn} \gamma t}  \mathcal{F}_{\cos \theta}(m-n)  + \frac{1}{2} \delta_{mn}.  \\
    \langle \hat c_m \hat c_n \rangle (t) &= e^{-\gamma |m-n| t} \langle \hat c_m \hat c_n \rangle(0) = -\frac{i}{2} e^{-\gamma |m-n| t} \mathcal{F}_{\sin \theta}(m-n).
\end{align}
From here, we can compute that 
\begin{align}
    n_k(t) &= \langle \hat \beta_k^\dagger \hat \beta_k \rangle(t) = \langle \cos(\theta_k) \hat d_k^\dagger \hat d_k + \sin^2(\theta_k/2) + i/2 \sin(\theta_k) (\hat d_k^\dagger \hat d_{-k}^\dagger - \hat d_{-k} \hat d_k) \rangle(t)  \nonumber \\
    &= \frac{1}{N} \sum_{mn} e^{-ik(m-n)} \left\langle \cos(\theta_k) \hat c_m^\dagger \hat c_n + \sin^2(\theta_k/2) + i/2 \sin(\theta_k) (\hat c_m^\dagger \hat c_n^\dagger - \hat c_m \hat c_n)\right\rangle(t)  \nonumber \\
    &= \sin^2(\theta_k/2) + \frac{1}{N} \sum_{mn} e^{-ik(m-n)} \cos(\theta_k) \langle \hat c_m^\dagger \hat c_n \rangle(t) - ie^{-ik(m-n)} \sin(\theta_k)  \langle \hat c_m \hat c_n \rangle(t)  \nonumber \\
    &= \frac{1}{2} - \frac{1}{2N} \sum_{mn} e^{-ik(m-n) - |m-n| \gamma t - \delta_{mn} \gamma t} \cos(\theta_k) \mathcal{F}_{\cos \theta}(m-n) + e^{-ik(m-n) - |m-n| \gamma t} \sin(\theta_k)  \mathcal{F}_{\sin \theta}(m-n)  \nonumber \\
    &= \frac{1}{2} - \frac{1}{2} e^{-\gamma t} \cos(\theta_k) \mathcal{F}_{\cos \theta}(0) -\frac{1}{2} \sum_{d \neq 0} e^{-ikd - |d| \gamma t } \cos(\theta_k) \mathcal{F}_{\cos \theta}(d) + e^{-ikd - |d| \gamma t} \sin(\theta_k)  \mathcal{F}_{\sin \theta}(d)  \nonumber \\
    &= \frac{1}{2} - \frac{1}{2} e^{-\gamma t} \cos(\theta_k) \mathcal{F}_{\cos \theta}(0) - \sum_{d =1}^\infty e^{- d \gamma t } \cos(k d) \cos(\theta_k) \mathcal{F}_{\cos \theta}(d) -i e^{ - d \gamma t} \sin(k d) \sin(\theta_k)  \mathcal{F}_{\sin \theta}(d).
\end{align}
This gives the general result for the quasiparticle number under the decoherence channel. If we restrict ourself to considering the quasiparticle number starting from the critical ground state, we can further simplify this expression. Using \cref{eqn:cor1,eqn:cor2}, we have that
\begin{align}
    n_k(t) &= \frac{1}{2} - \frac{1}{\pi} e^{-\gamma t} |\sin(k/2)|  - \frac{\mathrm{sgn}(k)}{\pi} \sum_{d =1}^\infty e^{- d \gamma t } \cos(k d) \sin(k/2) \frac{2}{1 - 4 d^2} - e^{ - d \gamma t} \sin(k d) \cos(k/2)  \frac{4 d}{1 - 4 d^2}  \nonumber \\
    &= \frac{1}{2} + \frac{1}{\pi}  |\sin(k/2)| (1 - e^{-\gamma t}) - \frac{i}{\pi} \cosh(\gamma t/2) \left[ \mathrm{arctanh}\left( e^{-(i|k| + \gamma t)/2} \right) - \mathrm{arctanh}\left( e^{(i|k| - \gamma t)/2} \right) \right]. \label{eqn:nkt}
\end{align}
This expression is in closed form, but we can get some more intuition by taking limiting forms. First, let's try and understand the expansion when $|k| \ll \gamma t$. Defining $x = |k|/(\gamma t)$ to be a small parameter and Taylor expanding gives us that
\begin{align}
    i\left[ \mathrm{arctanh}\left( e^{-(i|k| + \gamma t)/2} \right) - \mathrm{arctanh}\left( e^{(i|k| - \gamma t)/2} \right) \right] &= i\left[ \mathrm{arctanh}\left( e^{-\gamma t/2(-i x - 1)} \right) - \mathrm{arctanh}\left( e^{-\gamma t/2(i x - 1)}  \right) \right]  \nonumber \\
    &= \frac{\gamma t x}{2\sinh(\gamma t)} + \mathcal{O}(x^3) = \frac{|\sin(k/2)|}{\sinh(\gamma t)} + \mathcal{O}(x^3). \\
    \implies n_k(t) &= \frac{1}{2} + \frac{1}{\pi}  |\sin(k/2)| (1 - e^{-\gamma t} - \coth(t/2))  + \mathcal{O}(|k|/\gamma t)^3.
\end{align}
Asymptotically, this looks like exponential relaxation to the steady state value of $n_k = 1/2$, but starting from an initial condition $n_k(0) = 1/2 - 3|\sin(k/2)|/\pi = 1/2 - 3\epsilon_k/4\pi$. We can think of the dynamics as a quick ``jump'' at short times by an energy dependent amount, followed by slow exponential relaxation at a rate that is independent of energy.

We can similarly take the opposite limit, where we assume that $x = \gamma t/|k|$ is now a small parameter. In this case, we find that
\begin{align}
    i\left[ \mathrm{arctanh}\left( e^{-(i|k| + \gamma t)/2} \right) - \mathrm{arctanh}\left( e^{(i|k| - \gamma t)/2} \right) \right] &= i\left[ \mathrm{arctanh}\left( e^{|k|(-i - x)/2} \right) - \mathrm{arctanh}\left( e^{|k|(i - x)/2} \right) \right]  \nonumber \\
    &= -\frac{\pi}{2} - \frac{|k| x}{2 |\sin(k/2)|} + \mathcal{O}(x^3)  \nonumber \\
    &= -\frac{\pi}{2} - \frac{\gamma t}{2 |\sin(k/2)|} + \mathcal{O}(\gamma t/|k|)^3. \\
    \implies n_k(t) &= \left( |\sin(k/2)| + \frac{1}{2 |\sin(k/2)|} \right) \frac{\gamma t}{\pi} + \mathcal{O}(\gamma t/|k|)^3.
\end{align}
Given that we have taken $x = \gamma t/|k|$ to be a small fixed value, the dynamics for short times can be thought of as  linear growth at a rate $\gamma/(2 \pi |\sin(k/2)|)$.

\subsection{Effective temperature}
\label{app:B.2}

Using the quasiparticle expectations, we can define an effective non-equilibrium temperature as a function of energy and time. A temperature that is nearly independent of the energy demonstrates a near equilibrium distribution, whereas a temperature that depends strongly on the energy is a very far from equilibrium distribution. Generically, one would imagine that, because the noise process we have described is energy-agnostic (being infinite bandwidth and infinite temperature), that there is no reason that one would find an equilibrium distribution. However, we can use the limiting descriptions to try and describe an effective temperature in the model. We define the temperature using the Fermi-Dirac distribution:
\begin{align}
    n(\epsilon) &= \frac{1}{1 + e^{\epsilon/T}} \implies T_\mathrm{eff}(\epsilon,t) = \frac{\epsilon}{\log(1 - n) - \log(n)}.
\end{align}
At the phase transition, the energy is given by
\begin{align}
    \epsilon_k &= 2 \sqrt{(\cos k -1)^2 + \sin^2(k)} = 2 \sqrt{4 \sin^2(k/2)} = 4 |\sin(k/2)|.
\end{align}
We can then use this to calculate the temperature from the limiting forms of the quasiparticle number via
\begin{align}
    T_\mathrm{eff} &= \left\{ \begin{array}{cc}
        \pi  \frac{e^{2\gamma t} - e^{\gamma t}}{3 e^{\gamma t} - 1} + \mathcal{O}(|k|/\gamma t)^2 & |k| \ll \gamma t  \\
        -4\epsilon_k \log^{-1} \left[ \left( \frac{\epsilon_k}{2J} + \frac{J}{\epsilon_k} \right) \frac{\gamma t}{\pi}  \right] + \mathcal{O}(\gamma t/|k|)^2 & |k| \gg \gamma t
    \end{array} \right. ,
\end{align}
where, as described in the main text we have that once the time is longer than the wavevector --- or equivalently the effective length scale $\xi_F = 1/\gamma t$ is smaller than the effective wavelength $\lambda \sim 1/|k|$ --- then the distribution begins to look thermal with a temperature tending to infinity as a function of time. 

To understand how surprising it is that there is an emergent temperature, let's instead consider noise in the $z$-direction instead of the $x-y$ plane. One can again calculate the time evolution of the quasiparticle number analytically using 
\begin{align}
    \langle \hat c_m^\dagger \hat c_n \rangle(t) &= (1 - \delta_{mn})e^{-\gamma t} \langle \hat c_m^\dagger \hat c_n \rangle(0)+ \delta_{mn} \langle \hat c_m^\dagger \hat c_n \rangle(0) , \\
    \langle \hat c_m \hat c_n \rangle(t) &= e^{-\gamma t} \langle \hat c_m \hat c_n \rangle(0) ,
\end{align}
to find that (for anywhere in the phase space) we have
\begin{align}
    n_k(t) &= n_k(\infty) (1 - e^{-\gamma t}) , \\
    n_k(\infty) &= \frac{1}{2}\left[1 - \cos(\theta_k) \mathcal{F}_{\cos \theta}(0) \right] ,
\end{align}
which just gives uniform relaxation to a steady state value that depends weakly and non-monotonically on $k$. Working at the phase transition, this can be written as
\begin{align}
    n_k(\infty) &= \frac{1}{2}\left( 1 - \frac{2|\sin(k/2)|}{\pi} \right) = \frac{1}{2}\left( 1 - \frac{\epsilon_k}{2\pi} \right), 
\end{align}
which to leading order gives a temperature that depends linearly on the energy is extremely non-equilibrium:
\begin{align}
    T_\mathrm{eff}(\epsilon,t) &= \frac{\epsilon}{2 \mathrm{arccoth}(e^{\gamma t})} + \mathcal{O}(\epsilon^2).
\end{align}
A third noise model one could consider is neglecting the strings in the spin dissipation. This would be modeled by the free fermion model
\begin{align}
    \partial_t \hat \rho &= \gamma \sum_i \D[\hat c_i ] \hat \rho + \D[\hat c_i^\dagger ] \hat \rho
\end{align}
The dynamics induced by such a channel are significantly different from the $XY$ noise, and one can show that
\begin{align}
    n_k(t) &= \frac{1}{2}(1 - e^{-2 \gamma t}),
\end{align}
regardless of the value of $g$. Therefore, the temperature can be written exactly for all times as
\begin{align}
    T_\mathrm{eff}(\epsilon,t) &= \frac{\epsilon}{2 \mathrm{arccoth}(e^{2 \gamma t})},
\end{align}
with no corrections.

\subsection{Other correlation functions}
\label{app:B.3}

The only other two point functions allowed by translational invariance are of the form $C_k = \langle \hat \beta_k \hat \beta_{-k} \rangle$. These can also be computed exactly as a function of time:
\begin{align}
    C_k &= \left\langle  \cos^2(\theta_k/2) \hat d_k \hat d_{-k} + \sin^2(\theta_k/2) \hat d_{-k}^\dagger \hat d_{k}^\dagger  + i/2  \sin(\theta_k) [ \hat d_{-k}^\dagger \hat d_{-k} - \hat d_k \hat d_k^\dagger ] \right\rangle  \nonumber \\
    &= \frac{1}{N} \sum_{mn} e^{-ik(m-n)} \left\langle  \cos^2(\theta_k/2) \hat c_m \hat c_n + \sin^2(\theta_k/2) c_m^\dagger c_n^\dagger  + i \sin(\theta_k)  c_m^\dagger c_n  \right\rangle - i/2 \sin(\theta_k)  \nonumber \\
    &= \frac{1}{N} \sum_{mn} e^{-ik(m-n)} \left\langle  \cos(\theta_k) \hat c_m \hat c_n   + i \sin(\theta_k)  c_m^\dagger c_n  \right\rangle - i/2 \sin(\theta_k)   \nonumber \\
    &= -\frac{i}{2} \sum_{d} e^{-ik d - |d|\gamma t - \delta_{d,0}\gamma t} \left( \cos(\theta_k) \mathcal{F}_{\sin \theta}(d) - \sin(\theta_k) \mathcal{F}_{\cos \theta}(d) \right) .
\end{align}
Once more, we can evaluate this at the phase transition to find that
\begin{align}
    C_k &= -\frac{i \mathrm{sgn}(k)}{ \pi}  \sum_{d} e^{-ik d - |d|\gamma t - \delta_{d,0}\gamma t} \frac{-2id\sin(k/2) - \cos(k/2) }{1 - 4d^2}  \nonumber \\
    &= \frac{i \mathrm{sgn}(k)}{ \pi} \cos(k/2) e^{-\gamma t} + \frac{2i \mathrm{sgn}(k)}{ \pi}  \sum_{d=1}^\infty e^{- |d|\gamma t } \frac{2d\sin(k/2) \sin(kd) + \cos(k/2) \cos(kd) }{1 - 4d^2}  \nonumber \\
    &= \frac{i \mathrm{sgn}(k)}{ \pi} \cos(k/2)( e^{-\gamma t} - 1) + \frac{i \mathrm{sgn}(k)}{ \pi} \sinh(\gamma t/2) \left[ \mathrm{arctanh}\left( e^{-(i|k| + \gamma t)/2} \right) + \mathrm{arctanh}\left( e^{(i|k| - \gamma t)/2} \right) \right].
\end{align}
This function tends to be significantly smaller than the populations $n_k$. It is largest for $k = 0$, where
\begin{align}
    C_0 &\to \frac{i}{\pi} \left[ e^{-\gamma t} - 1  + 2\sinh(\gamma t/2) \mathrm{arctanh} \left( e^{-\gamma t/2} \right) \right].
\end{align}
The maximal value of this is given by the solution of a transcendental equation $\partial_t C_0(t)|_{t = t^*} = 0$. The solution found numerically is $\gamma t^* = 0.41242...$, $C_0(t^*) = 0.04285...$. Hence, the anomalous correlation functions are always at least an order of magnitude smaller than the densities. Similarly to the local density, this correlator grows superlinearly at short times:
\begin{align}
    C_0(t) &= -\frac{i}{2 \pi} \gamma t \log \gamma t + \mathcal{O}(t).
\end{align}

We can also calculate analytically the anomalous correlators for the other noise models. For the free fermion noise, we have that
\begin{align}
    C^\mathrm{ff}_k(t) &= e^{-\gamma t} C^\mathrm{ff}_k(0) = 0.
\end{align}
Since the initial condition has no coherences, this is just identically zero for all times. For $Z$ noise, it is useful to calculate the time dynamics of the plane wave modes as an intermediate step. 
\begin{align}
    \langle \hat d_k^\dagger \hat d_k \rangle(t) &= \frac{1}{N} \sum_{mn} e^{ik(m-n)} \langle \hat c_m^\dagger \hat c_n \rangle(t) = \frac{1}{N} \sum_{mn} e^{ik(m-n)} e^{-\gamma t} \langle \hat c_m^\dagger \hat c_n \rangle(0) + \frac{1}{N} \sum_{mn} e^{ik(m-n)} \delta_{mn} (1 - e^{-\gamma t})\langle \hat c_m^\dagger \hat c_n \rangle(0) \nonumber \\
    &= e^{-\gamma t} \langle \hat d_k^\dagger \hat d_k \rangle(0) + (1 - e^{-\gamma t}) \overline{n}, \\
    \langle \hat d_{-k} \hat d_k \rangle(t) &= e^{-\gamma t} \langle \hat d_{-k} \hat d_k \rangle(0),
\end{align}
where $\overline{n}$ is the mean particle density (or magnetization) in the initial condition, which is unchanged by the noise. From here, it is simple to calculate that the time evolution of the coherences is given by
\begin{align}
    C_k^Z(t) &= \left\langle  \cos^2(\theta_k/2) \hat d_k \hat d_{-k} + \sin^2(\theta_k/2) \hat d_{-k}^\dagger \hat d_{k}^\dagger  + i/2  \sin(\theta_k) [ \hat d_{-k}^\dagger \hat d_{-k} - \hat d_k \hat d_k^\dagger ] \right\rangle(t)  \nonumber \\
    &= e^{-\gamma t} C^Z_k(0) + \frac{i}{2}(1 - e^{-\gamma t}) \sin(\theta_k)(2 \overline{n} - 1) .
\end{align}
Working at the phase transition and taking $C_k(0) = 0$, we have $\overline{n} = \frac{1}{2} - \frac{1}{\pi}$ so
\begin{align}
    C_k^Z(t) &= -\frac{i}{\pi} \mathrm{sgn}(k) \cos(k/2) (1 - e^{-\gamma t}) .
\end{align}

\subsection{Divergent quasiparticle production rate}
\label{app:B.4}

In addition to the energy resolved QP numbers, we can also calculate the total QP number via
\begin{align}
    n(t) &= \int \frac{\dd k}{2 \pi} n_k(t) = \frac{1}{2} - \frac{1}{2} |\mathcal{F}_{\cos \theta}(0)|^2 e^{-\gamma t} - \sum_{d = 1}^\infty e^{-d \gamma t} \left( |\mathcal{F}_{\cos \theta}(d)|^2 + |\mathcal{F}_{\sin \theta}(d)|^2\right),
\end{align}
at the phase transition, this can be written as
\begin{align}
    n(t) &= \int \frac{\dd k}{2 \pi} n_k(t) = \frac{1}{2} - \frac{2}{\pi^2}  e^{-\gamma t} - \frac{4}{\pi^2} \sum_{d = 1}^\infty e^{-d \gamma t} \frac{1 + 4d^2}{(1 - 4d^2)^2} . \label{eqn:nt}
\end{align}
We can note that the sum is convergent for all $t \geq 0$, as expected. This sum can be rewritten in terms of special functions, but these are not necessarily intuitive. To try and understand the short time dynamics, let's first note that when taking a derivative, the sum will be divergent, so we cannot naively perform a short time expansion. Hence, observe that 
\begin{align}
    \sum_{d = 1}^\infty e^{-d \gamma t} \frac{1 + 4d^2}{(1 - 4d^2)^2} &= - \sum_{d = 1}^\infty  \frac{e^{-d \gamma t}}{(1 - 4d^2)} + \sum_{d = 1}^\infty  \frac{2 e^{-d \gamma t}}{(1 - 4d^2)^2} \nonumber \\
    &= \frac{1}{2} - \sinh(\gamma t/2) \mathrm{arccoth}(e^{\gamma t/2}) + \sum_{d = 1}^\infty  \frac{2 e^{-d \gamma t}}{(1 - 4d^2)^2}.
\end{align}
The second sum is still difficult to evaluate, but the leading order terms are now convergent, and so we can observe that
\begin{align}
    \sum_{d = 1}^\infty e^{-d \gamma t} \frac{1 + 4d^2}{(1 - 4d^2)^2} &= \frac{1}{2} -           \sinh(\gamma t/2) \mathrm{arccoth}(e^{\gamma t/2}) + \frac{\pi^2 -8}{8} - \frac{\gamma t}{8} + \mathcal{O}(\gamma t)^2, \\
    \implies n(t) &= \frac{4}{\pi^2} \sinh(\gamma t/2) \mathrm{arccoth}(e^{\gamma t/2}) + \frac{5}{2\pi^2} \gamma t + \mathcal{O}(\gamma t)^2 \nonumber \\
    &= -\frac{1}{\pi^2} \gamma t \log(\gamma t) + \frac{5 + 4 \log 2}{2 \pi^2} \gamma t + \mathcal{O}(\gamma t)^2 .
\end{align}
From here, we can take a derivative to find that
\begin{align}
    \partial_t n(t) &= -\frac{\gamma}{\pi^2}  \log(\gamma t) + \frac{3 + 4 \log 2}{2 \pi^2} \gamma  + \mathcal{O}(\gamma t) .
\end{align}
which has a logarithmic divergence for small $t$. For completeness, we also note that we can define the long time behavior by noting that \cref{eqn:nt} is a power series expansion in $e^{-\gamma t}$. Keeping the lowest order terms gives
\begin{align}
    n(t) &= \frac{1}{2} - \frac{38}{9 \pi^2} e^{-\gamma t} + \mathcal{O}(e^{-\gamma t})^2
\end{align}

\begin{figure}
    \centering
    \includegraphics[width=0.5\linewidth]{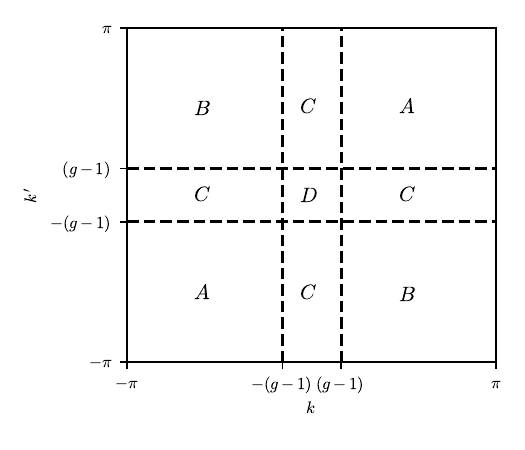}
    \caption{Sketch depicting the four symmetry-distinct regions of the integral in \cref{eqn:integral}.}
    \label{fig:intFig}
\end{figure}

Going back to the short time expansion, the derivative of $n(t)$ should only be divergent when the state is critical. We can understand how this divergence appears near the phase transition by writing directly the time derivative in the number of quasiparticles at time $t = 0$:
\begin{align}
    \partial_t n|_{t = 0} &=  \frac{\gamma}{2} |\mathcal{F}_{\cos \theta}(0)|^2  + \sum_{d = 1}^\infty  d \gamma \left( |\mathcal{F}_{\cos \theta}(d)|^2 + |\mathcal{F}_{\sin \theta}(d)|^2\right) \nonumber \\
    &\sim \frac{\gamma}{2} |\mathcal{F}_{\cos \theta}(0)|^2 -2 \gamma \int_{-\pi}^\pi \frac{\dd k}{2 \pi} \frac{\dd k'}{2 \pi} \int_0^\infty \dd x  e^{ix(k-k')} x \sin^2([\theta_k - \theta_{k'}]/2) \nonumber \\
    &= \frac{\gamma}{2} |\mathcal{F}_{\cos \theta}(0)|^2 + 2 \gamma  \int_{-\pi}^\pi \frac{\dd k}{2 \pi} \frac{\dd k'}{2 \pi} \frac{\sin^2([\theta_k - \theta_{k'}]/2)}{(k-k')^2}. \label{eqn:integral}
\end{align}
We wish to evaluate this just to leading order in $g-1$. To perform this integral, it will be useful to subdivide this into 9 relevant regions (though only four independent ones after considering symmetries), as depicted in \cref{fig:intFig}. Let's define the integrand $f_g(k,k') = \frac{1}{4 \pi^2} \frac{\sin^2([\theta_k - \theta_{k'}]/2)}{(k-k')^2}$:
\begin{align}
    \int_{[-\pi, \pi] \times [-\pi,\pi]} f_g(k,k') \dd k \dd k' &= 2\int_A f_g(k,k') \dd k \dd k' + 2\int_B f_g(k,k') \dd k \dd k' + 4\int_C f_g(k,k') \dd k \dd k' + \int_D f_g(k,k') \dd k \dd k'
\end{align}
We will evaluate each one separately to leading order. Let's begin with $C$, which is defined as $g-1 < k < \pi$ and $-(g-1) < k' < g-1$.  Then we can note that $0 \leq f_g \leq 1/(2 \pi k)^2$. This implies that
\begin{align}
    \int_C f_g(k,k') \dd k \dd k' \leq \int_C \frac{1}{(2 \pi k)^2} \dd k \dd k'   = \frac{g-1}{2 \pi^2} \left( \frac{1}{g-1} - \frac{1}{\pi} \right) \leq \frac{1}{2 \pi^2} = \mathcal{O}(g-1)^0, 
\end{align}
Next, let's evaluate $D$ where $0 < |k|,|k'| < g-1$. Here, we have that $f_g \leq 1/[16 \pi^2 (g-1)^2]$. Therefore, $\int_D f_g(k,k') \dd k \dd k' \leq 1/(16 \pi^2) = \mathcal{O}(g-1)^0$ as well. To evaluate $A$ and $B$, we first need to find the leading order behavior $f_g$ in $g-1$. We have that
\begin{align}
    f_g(k,k') &= f_1(k,k') - \frac{\sin^2[(k-k')/2]}{16 \pi^2 (k-k')^2\sin(k/2) \sin(k'/2)}(g-1) \equiv f_1(k,k') - \tilde f(k,k')(g-1).
\end{align}
Here, $f_1$ is the function $f_g$ evaluated at $g = 1$, and we have defined $\tilde f$ to be the leading correction. Next, we note that 
\begin{align}
    |\tilde f| &\leq \frac{1}{64 \pi^2} \frac{1}{\sin(k/2) \sin(k'/2)}, \\
    \implies \left| \int_{g-1}^\pi \int_{g-1}^\pi \tilde f \right| &\leq \frac{1}{64 \pi^2} \left| \int_{g-1}^\pi \frac{1}{\sin(k/2)} \dd k \right|^2  \leq \frac{1}{64 \pi^2} \left| \int_{g-1}^\pi \frac{\pi}{k} \dd k \right|^2 = \frac{1}{64} \left| \log(\pi) - \log(g-1) \right|^2, 
\end{align}
and so the correction is of order $[\log^2(g-1)] (g-1) \to 0$ as $g \to 1$, and is therefore smaller than $\mathcal{O}(g-1)^0$. Hence, we can evaluate the $A$ and $B$ using the values of $\theta_k$ for $g = 1$. This significantly simplifies the problem, allowing us to write that
\begin{align}
    \int_A \frac{\sin([k-k']/4)^2}{(k-k')^2} \frac{\dd k \dd k'}{4 \pi^2} \leq \int_{g-1}^\pi \int_{g-1}^\pi \frac{1}{16} \frac{\dd k \dd k'}{4 \pi^2} \leq \frac{1}{64} = \mathcal{O}(g-1)^0, 
\end{align}
Finally, we come to evaluating $B$, which is given by
\begin{align}
    \int_B \frac{\cos([k-k']/4)^2}{(k-k')^2} \frac{\dd k \dd k'}{4 \pi^2} =  \int_{g-1}^\pi \int_{g-1}^\pi \frac{1}{(k+k')^2} \frac{\dd k \dd k'}{4 \pi^2} -  \int_{g-1}^\pi \int_{g-1}^\pi \frac{\sin([k+k']/4)^2}{(k+k')^2} \frac{\dd k \dd k'}{4 \pi^2}.
\end{align}
We only want the divergent part, and from before we know the second term is bounded by a constant. Hence, we care about the first piece:
\begin{align}
     \int_{g-1}^\pi \int_{g-1}^\pi \frac{1}{(k+k')^2} \frac{\dd k \dd k'}{4 \pi^2} = \frac{1}{4 \pi^2} \left[ 2 \log(g -1 + \pi) - \log(4 \pi) - \log(g-1) \right].
\end{align}
Putting everything together, we have that
\begin{align}
    \partial_t n\vert_{t = 0} &= -\frac{\gamma}{\pi^2} \log(g-1) + \mathcal{O}(g-1)^0.
\end{align}

Another way to see the logarthmic divergence is by using \cref{eqn:nkt}. It is incredibly simple to calculate the energy resolved QP production rate:
\begin{align}
     \partial_t n_k &= \frac{\gamma |\sin(k/2)|}{\pi} e^{-\gamma t} + \frac{\gamma }{2 \pi} \left( \frac{\cosh(\gamma t/2)^2 |\sin(k/2)|}{|\sin[(k - i \gamma t)/2]|^2} - i \sinh(\gamma t/2) \left[ \mathrm{arctanh} \left(e^{(ik + \gamma t)/2} \right) - \mathrm{arctanh} \left(e^{(-ik + \gamma t)/2} \right) \right] \right). \\
   &  \implies  \partial_t n_k|_{t = 0} = \frac{\gamma}{\pi} |\sin(k/2)| + \frac{\gamma}{2\pi} \frac{1}{|\sin(k/2)|} = \frac{2\gamma}{\pi} \left(  \frac{J}{\epsilon_k} + \frac{\epsilon_k}{J 8}\right).
\end{align}
If we note that the gap is $|g-1|$ and assume the derivatives are essentially unchanged near the critical point, then we can use the fact that the density of states near the critical point is $\rho(\epsilon) = 1/2\pi$ to calculate that
\begin{align}
    \partial_t n &= \int_{|g-1|}^2 \dd \epsilon \rho(\epsilon) \partial_t n_\epsilon \sim \int_{|g-1|}^2 \frac{\gamma}{\pi^2} \frac{1}{\epsilon} \dd \epsilon  \sim -\frac{\gamma}{\pi^2} \log(g-1),
\end{align}
where $\sim$ means we are only keep leading order terms.

\section{Critical XX Model}
\label{app:C}

In this section, we will show that the basic phenomena found for the critical Ising model also extends to a critical XX model. Here will consider the Hamiltonian given by
\begin{align}
    \hat H &= -\frac{J}{2}\sum_{i = 1}^{N-1} \hat \sigma_i^x \hat \sigma_{i + 1}^x + \hat \sigma_i^y \hat \sigma_{i + 1}^y = -J\sum_{i = 1}^{N-1} \hat c_i^\dagger \hat c_{i + 1} + \hat c_{i + 1}^\dagger \hat c_i.
\end{align}
Written in terms of the plane wave momentum modes $\hat d_k$, this is diagonalized:
\begin{align}
    \hat H &= -2J \sum_k \cos(k) \hat d_k^\dagger \hat d_k.
\end{align}
We will consider this model in the zero magnetization sector, which in its ground state is given by 
\begin{align}
    \langle \hat d_k^\dagger \hat d_k \rangle_\mathrm{GS} &= \left\{ \begin{array}{cc}
       1  & |k| < \pi/2  \\
       0  & |k| > \pi/2
    \end{array} \right., \\
    \implies \langle \hat c_m^\dagger \hat c_n \rangle_\mathrm{GS} &= \int_{-\pi/2}^{\pi/2} e^{ik(m-n)} \frac{\dd k}{2 \pi} = \frac{\sin \left( \frac{\pi}{2} [m-n]\right)}{\pi(m-n)}.
\end{align}
From here, we can use the already known solution observe that the time evolution is given by
\begin{align}
    \langle \hat c_m^\dagger \hat c_n \rangle(t) &= (1 - \delta_{mn}) e^{-|m-n|\gamma t} \langle \hat c_m^\dagger \hat c_n \rangle_{\mathrm{GS}} +  \delta_{mn} \langle \hat c_m^\dagger \hat c_n \rangle_{\mathrm{GS}} , \\
    \implies \langle \hat d_k^\dagger \hat d_k \rangle(t) &= \frac{1}{N}\sum_{m,n} e^{-ik(m-n)}  \langle \hat c_m^\dagger \hat c_n \rangle(t) \nonumber \\
    &= \frac{1}{N}\sum_{m,n} e^{-ik(m-n)}  \left[ (1 - \delta_{mn}) e^{-|m-n|\gamma  t} \langle \hat c_m^\dagger \hat c_n \rangle_{\mathrm{GS}} +  \delta_{mn} \langle \hat c_m^\dagger \hat c_n \rangle_{\mathrm{GS}} \right] \nonumber \\
    &= \frac{1}{2} + 2\sum_{d > 0} e^{ - \gamma |d|t} \cos(k d) \frac{\sin(\pi d/2)}{\pi d} \nonumber  \\
    &= \frac{1}{2} + \frac{\arccot(e^{-i k + \gamma  t}) + \arccot (e^{i k + \gamma  t})}{\pi}.
\end{align}
From here, we have all the quasiparticle operators defined for all times. We can calculate the long and short time expansions
\begin{align}
    n_k(t) &= \left\{ \begin{array}{cc}
       n_k(0) + \frac{2 J}{\pi \epsilon_k}\gamma t & \gamma t \lesssim |k|  \\
        \frac{1}{2} - \frac{\epsilon_k}{J \pi} e^{-\gamma t} & \gamma t \gtrsim |k| 
    \end{array} \right. .
\end{align}

We can again directly define the effective temperature as
\begin{align}
    T_\mathrm{eff}(k,t) &= \frac{-2J \cos(k)}{\log(1 - n_k) - \log(n_k)} . \label{eqn:temp_XX}
\end{align}
To understand the emergent temperature at long times, let's define a small parameter $z = (k - \pi/2)/t$. Here, we work around the $k = \pi/2$ point as this is where the system is gapless (similarly, the $k = -\pi/2$) points but inversion symmetry implies this will behave identically. Hence, we can calculate that
\begin{align}
    T_{\mathrm{eff}}(k,t) &= \frac{\pi}{2} J \sinh(\gamma   t) + \mathcal{O}(z^2) . \label{eqn:temp_approx_XX}
\end{align}
Similarly, we can take the short time expansion to get that
\begin{align}
    T_{\mathrm{eff}}(k,t) &= |\epsilon_k| \log^{-1} \left( \frac{\pi |\epsilon_k|}{2 J \gamma t} \right) + \mathcal{O}(z^{-2}) .\label{eqn:temp2_approx_XX}
\end{align}
Both of these expansions, up to constant factors, have essentially the same functional forms as was observed in the TFIM.

We can similarly calculate how the total quasiparticle number grows at short times. Now, Because we start in the zero magnetization sector (which corresponds to half filling in the fermionic language), the total particle number does not change. However, we can instead calculate the imbalance between positive and negative energies. For this, we note that the density of positive energy quasiparticles is given by
\begin{align}
    n_{\epsilon > 0} &= \int_{\pi/2}^\pi \langle \hat d_k^\dagger \hat d_k \rangle(t) \frac{\dd k}{\pi/2} = \frac{2}{\pi} \int_{\pi/2}^\pi \langle \hat d_k^\dagger \hat d_k \rangle(t) \dd k = \frac{1}{2} - \frac{2}{\pi^2} \left( \mathrm{Li}_2(e^{-\gamma t}) - \mathrm{Li}_2(-e^{-\gamma t}) \right) , \\
    \implies n_{\epsilon > 0} &= -\frac{2}{\pi^2}\gamma t \log(\gamma t) + \frac{2}{\pi^2}(1 + \log(2)) \gamma t + \mathcal{O}(\gamma t)^2 ,\\
    \implies n_{\epsilon > 0} &= \frac{1}{2} -\frac{4}{\pi^2} e^{-\gamma t} + \mathcal{O}(e^{-\gamma t})^3
\end{align}
where $\mathrm{Li}_2(z)$ is the dilogarithm function. Hence, the short time particles number grows superlinearly, and the long time relaxation looks as if it started with an initial condition of $(1/2 - 4/\pi^2)$, just as in the TFIM. We can take a derivative of this to find that
\begin{align}
    \partial_t n_{\epsilon > 0} &= \frac{2 \gamma}{\pi^2} \log \left[ \coth \left( \frac{\gamma t}{2} \right) \right] = - \frac{2 \gamma }{\pi^2} \log(\gamma t) + \frac{2 \gamma }{\pi^2} \log(2) + \mathcal{O}(t^2) ,
\end{align}
and so we see an almost identical logarthmic divergence as in the Ising model, up to multiplicative prefactors.

Finally, we can exactly calculate the energy resolved quasiparticle production. This is rather simple:
\begin{align}
    &\partial_t \langle \hat d_k^\dagger \hat d_k \rangle = -\frac{2 \gamma }{\pi} \frac{\cos(k) \cosh(\gamma t)}{\cos(2 k) + \cosh(2 t)} = \frac{2 \gamma }{\pi} \frac{J \epsilon_k \cosh(\gamma t)}{\epsilon_k^2 + 2 J^2 \sinh(\gamma t)} , \\
    \implies &\partial_t \langle \hat d_k^\dagger \hat d_k \rangle \big\vert_{t = 0} = \frac{2 \gamma  J}{\pi \epsilon_k} ,
\end{align}
which is identical to the result for the TFIM to leading order in $\epsilon_k$.

\section{Experimental Protocols}
\label{app:D}

\subsection{Measuring the effective temperature}
\label{app:D.1}

Here, we propose a possible method to measure the effective temperature of the many-body state using a single qubit probe. The proposed measurement makes use of the fact that a weakly coupled probe qubit to one end of the chain can probe the frequency-dependent density of states of the many-body state. We imagine the following sequence:
\begin{enumerate}
    \item Prepare the many-body ground state of the Ising model.
    \item Apply the noise channel for a time $t$.
    \item Turn the noise channel off, and the Ising Hamiltonian on. The populations of eigenmodes are invariant under the Hamiltonian evolution.
    \item Weakly couple a tunable qubit using an XX interaction to one end of the spin chain with an energy gap $\Delta$ beween $\hat \sigma^z$ eigenstates.
    \item The qubit will equilibrate to a mixture of its energy eigenstates. By measuring the relative population of the spin up and spin down states, one can define a temperature as a function of $\Delta$.
\end{enumerate}
What remains to be shown is that the equilibrium distribution of the weakly coupled probe qubit will have the same effective temperature as the many-body spin state. To see this, we can calculate the effective heating and cooling rates of the probe qubit, which we will denote via the operators $\hat \sigma_P^{x,y,z}$. We imagine a Hamiltonian 
\begin{align}
    \hat H &= \hat H_\mathrm{Ising} + \frac{\Delta}{2} \hat \sigma_P^z + \lambda(\hat \sigma_P^+ \hat \sigma_1^- + \hat \sigma_P^- \hat \sigma_1^+), \\
    \hat H_\mathrm{Ising} &= -J\sum_{j = 1}^{N-1} \hat \sigma_j^x \hat \sigma_{j + 1}^x + g \sum_{j = 1}^N \hat \sigma_j^z,
\end{align}
where the Ising model is tuned to the critical point $g = J$, the intersystem coupling $\lambda \ll g,J$ and $\Delta$ is a tunable parameter to select which frequency in the Ising chain we are sensitive to. Note that, because the probe qubit is coupled to the end of the chain, the entire system is still free-fermion. I.e., using the JW transformation we can write the dynamics as
\begin{align}
    \hat H &= \hat H_\mathrm{Ising} + \Delta \hat c_P^\dagger \hat c_P + \lambda(\hat c_P^\dagger \hat c_1 + \hat c_1^\dagger \hat c_P), \\
    \hat H_\mathrm{Ising} &= -J \sum_{j = 1}^{N-1}  (\hat c_j^\dagger - \hat c_j ) (\hat c_{j + 1}^\dagger + \hat c_{j + 1}) +  2g \sum_{j = 1}^N \hat c_j^\dagger \hat c_j,
\end{align}
up to irrelevant constants. At this point, it will be useful to write things in terms of the energy eigenmodes of the Ising Hamiltonian:
\begin{align}
    \hat H &=  \Delta \hat c_P^\dagger \hat c_P + \int \frac{\dd k}{2\pi} \epsilon_k \hat \beta_k^\dagger \hat \beta_k +  \lambda \int \frac{\dd k}{2 \pi} \left[ \hat c_P^\dagger e^{ik} (\cos(\theta_k/2) \hat \beta_k - i \sin(\theta_k/2) \hat \beta_{-k}^\dagger )  + \hc  \right]. 
\end{align}
If we define $\hat n_P = \hat c_P^\dagger \hat c_P$ and work in the interaction picture, then we can observe that
\begin{align}
    \hat H_\mathrm{int} &= \lambda \int \frac{\dd k}{2 \pi} \left[ \cos(\theta_k/2) e^{ik + i(\Delta - \epsilon_k) t } \hat c_P^\dagger \hat \beta_k + \sin(\theta_k/2) e^{ik + i(\Delta + \epsilon_k) t } \hat c_P^\dagger \hat \beta_{-k}^\dagger + \hc  \right] .
\end{align}
Note that $\epsilon_k > 0$ and we take WLOG $\Delta > 0$, so only the term proportional to $\cos(\theta_k/2)$ can be resonant. We will make the rotating wave approximation and drop the non-resonant terms, as in the limit $\lambda \to 0, t \to \infty$ the off-resonant terms do not cause transitions.
\begin{align}
    \hat H_\mathrm{int,RWA} &= \lambda \int \frac{\dd k}{2 \pi} \left[ \cos(\theta_k/2) e^{ik + i(\Delta - \epsilon_k) t } \hat c_P^\dagger \hat \beta_k  + \hc  \right] .
\end{align}
From this expression, it is simple to read off the Fermi's Golden rule transition rates as
\begin{align}
    \Gamma_\uparrow &= \frac{\lambda^2}{2J} \cos^2(\theta_k/2) \langle \hat \beta_k^\dagger \hat \beta_k \rangle = \frac{\lambda^2}{2J} \cos^2(\theta_k/2) n_k, \\
    \Gamma_\downarrow &= \frac{\lambda^2}{2J} \cos^2(\theta_k/2) \langle \hat \beta_k \hat \beta_k^\dagger \rangle = \frac{\lambda^2}{2J} \cos^2(\theta_k/2) (1 - n_k),
\end{align}
where $k$ is such that $\epsilon_k = \Delta$. We assume we are studying the low-energy spectrum of the model, and the factor of $2J$ in the denominator is the local density of states at low energy, where we use that $\epsilon_k \sim 2 J |k| \implies \partial_k \epsilon = 2J$ so $\rho(\epsilon) = 1/(2J)$. In general, the density of states is given by $\rho(\Delta) = (2 J \sqrt{1 - \Delta^2/4})^{-1}$. From here, we find that
\begin{align}
    \partial_t n_P &= \Gamma_\uparrow (1 - n_P) - \Gamma_\downarrow n_P, \\ 
    \implies n_P(t) &= \frac{\Gamma_\uparrow}{\Gamma_\uparrow + \Gamma_\downarrow} (1 - e^{-(\Gamma_\uparrow + \Gamma_{\downarrow})t}).
\end{align}
This gives a steady state value and overall relaxation rate
\begin{align}
    n_P(t \to \infty) &= \frac{\Gamma_\uparrow}{\Gamma_\uparrow + \Gamma_\downarrow} = n_k, \\
    \Gamma_{\mathrm{eff}} &= \Gamma_\uparrow + \Gamma_\downarrow = \frac{\lambda^2}{2J} \cos^2(\theta_k/2),
\end{align}
and so the steady state spin population will be $\langle \hat \sigma_P^z \rangle = 2 n_P - 1 = 2n_k - 1$ where $n_k$ is the QP density in the Ising model at energy $\epsilon_k  =\Delta$. Hence, we can directly read the (non-)equilibrium temperature of the Ising model from the steady state distribution function of the probe spin. Indeed, if one were to assign a temperature to the probe spin assuming it had reached thermal equilibrium with the Ising spin bath, it would be the exact same effective temperature as the Ising spin model, exactly as desired.

We believe that this protocol could be extremely well suited for many experimental platforms. One particular example would be long Josephson-Junction (JJ) arrays, which have recently been suggested as a simulation platform to observe Ising criticality \cite{Bell2018,Roy2023,Maffi2024}. These would be particularly appealing as JJ arrays are able to reach extremely large system sizes and would be able to easily observe the bulk physics. Further, it would be relatively straightforward to add a superconducting qubit coupled to the end of the chain to perform the measurement protocol.

\subsection{Measuring the total quasiparticle number}
\label{app:D.2}

The previously outlined protocol gives the full spectrum of quasiparticle density as a function of their energy. However, if one were instead just interested in measuring the total quasiparticle number (e.g., to observe the logarithmic divergence), then a simpler scheme can be imagined. Here, we imagine that it is possible to completely adiabatically tune the parameter $g$ to zero. In this case, the Hamiltonian is simply given by the pure Ising model in 1D:
\begin{align}
    \hat H &= -J\sum_i \hat \sigma_i^x \hat \sigma_{i  +1}^x.
\end{align}
This Hamiltonian has a flat quasiparticle spectrum corresponding to domain walls, which each carry energy $2J$. Hence, by projecting each lattice site into an $\hat X$ eigenstate and then counting domain walls, one can count exactly the total number of quasiparticles in the system. Hence, if we assume that the sweep is truly adiabatic, then the number of QP excitations will not change and so they can be counted easily using an extremely simple projective measurement, assuming the lattice has single site readout and the ability to tune deep into the ferromagnetic regime. Alternatively, one can adiabatically sweep to the paramagnet defined by the Hamiltonian
\begin{align}
    \hat H &= -g \sum_i \hat \sigma_i^z,
\end{align}
which also has a flat spectrum. By measuring each spin now in the $\hat Z$ eigenstate basis allows one to measure the total quasiparticle number in the exact same way. 

\end{document}